\documentclass[modern]{aastex631}

\shorttitle{Photometry of Didymos in support of DART}
\shortauthors{Moskovitz et al.}
\graphicspath{{./}{figures/}}
\begin{document}

\title{Photometry of the Didymos system across the DART impact apparition}

\correspondingauthor{Nicholas Moskovitz}
\email{nmosko@lowell.edu}

\author[0000-0001-6765-6336]{Nicholas Moskovitz}
\affiliation{Lowell Observatory, 1400 West Mars Hill Road, Flagstaff, AZ 86004, USA}

\author[0000-0003-3091-5757]{Cristina Thomas}
\affiliation{Northern Arizona University}

\author[0000-0001-8434-9776]{Petr Pravec}
\affiliation{Astronomical Institute of the Academy of Sciences of the Czech Republic, Fri\v{c}ova 298, Ond\v{r}ejov, CZ-25165, Czech Republic}

\author[0000-0002-3818-7769]{Tim Lister}
\affiliation{Las Cumbres Observatory}

\author{Tom Polakis}
\affiliation{Lowell Observatory, 1400 West Mars Hill Road, Flagstaff, AZ 86004, USA}

\author[0000-0003-0412-9664]{David Osip}
\affiliation{Las Campanas Observatory, Chile}

\author[0000-0003-1008-7499]{Theodore Kareta}
\affiliation{Lowell Observatory, 1400 West Mars Hill Road, Flagstaff, AZ 86004, USA}

\author[0000-0003-2341-2238]{Agata Ro{\.z}ek}
\affiliation{Institute for Astronomy, University of Edinburgh, Royal Observatory, Edinburgh, EH9 3HJ, UK}

\author[0000-0003-3240-6497]{Steven R. Chesley}
\affiliation{Jet Propulsion Laboratory, California Institute of Technology \\
California, USA}

\author[0000-0003-4439-7014]{Shantanu P. Naidu}
\affiliation{Jet Propulsion Laboratory, California Institute of Technology \\
California, USA}

\author{Peter Scheirich}
\affiliation{Astronomical Institute of the Academy of Sciences of the Czech Republic, Fri\v{c}ova 298, Ond\v{r}ejov, CZ-25165, Czech Republic}

\author{William Ryan}
\affiliation{New Mexico Institute of Mining and Technology/Magdalena Ridge Observatory, 801 Leroy Place, Socorro, NM 87801}

\author{Eileen Ryan}
\affiliation{New Mexico Institute of Mining and Technology/Magdalena Ridge Observatory, 801 Leroy Place, Socorro, NM 87801}

\author[0000-0001-5306-6220]{Brian Skiff}
\affiliation{Lowell Observatory, 1400 West Mars Hill Road, Flagstaff, AZ 86004, USA}

\author[0000-0001-9328-2905]{Colin Snodgrass}
\affiliation{Institute for Astronomy, University of Edinburgh, Royal Observatory, Edinburgh, EH9 3HJ, UK}

\author[0000-0003-2781-6897]{Matthew M. Knight}
\affiliation{Physics Department, United States Naval Academy, 572C Holloway Rd, Annapolis, MD 21402, USA}

\author{Andrew S. Rivkin}
\affiliation{Johns Hopkins University Applied Physics Laboratory}

\author{Nancy L. Chabot}
\affiliation{Johns Hopkins University Applied Physics Laboratory}

\author[0000-0001-5549-5037]{Vova Ayvazian}
\affiliation{E. Kharadze Georgian National Astrophysical Observatory, Abastumani, Georgia}

\author{Irina Belskaya}
\affiliation{ LESIA, Observatoire de Paris, Université PSL, CNRS, Université Paris Cité, Sorbonne Université, Meudon, France}
\affiliation{Institute of Astronomy, V.N. Karazin Kharkiv National University, Kharkiv, Ukraine}

\author[0000-0001-6285-9847]{Zouhair Benkhaldoun}
\affiliation{Ouka\"{i}meden Observatory, High Energy Physics and Astrophysics Laboratory, Cadi Ayyad University, BP 2390, Marrakech, Morocco}

\author{Daniel N. Berte\c{s}teanu}
\affiliation{Astronomical Institute of the Romanian Academy}

\author[0000-0002-7520-8389]{Mariangela Bonavita}
\affiliation{Institute for Astronomy, University of Edinburgh, Royal Observatory, Edinburgh, EH9 3HJ, UK}

\author{Terrence H. Bressi}
\affiliation{LPL/UA}

\author[0000-0002-2079-179X]{Melissa J. Brucker}
\affiliation{LPL/UA}

\author[0000-0002-5854-4217]{Martin J. Burgdorf}
\affiliation{Universit{\"a}t Hamburg,  Faculty of Mathematics, Informatics and Natural Sciences, Department of Earth Sciences,  Meteorological Institute, Bundesstra\ss{}e 55,  20146 Hamburg, Germany}

\author[0000-0003-1169-6763]{Otabek Burkhonov}
\affiliation{Ulugh Beg Astronomical Institute, Tashkent, Uzbekistan}

\author[0000-0002-6423-0716]{Brian Burt}
\affiliation{Lowell Observatory, 1400 West Mars Hill Road, Flagstaff, AZ 86004, USA}

\author[0000-0001-6293-9062]{Carlos Contreras}
\affiliation{Las Campanas Observatory, Chile}

\author[0000-0002-1278-5998]{Joseph Chatelain}
\affiliation{Las Cumbres Observatory}

\author[0000-0001-6060-5851]{Young-Jun Choi}
\affiliation{Korea Astronomy and Space Science Institute, 776, Daedeokdae-ro, Yuseong-gu, Daejeon 34055, Korea}
\affiliation{University of Science and Technology, 217, Gajeong-ro, Yuseong-gu, Daejeon 34113, Korea}

\author{Matthew Daily} 
\affiliation{Las Cumbres Observatory}

\author{Julia de Le\'on}
\affiliation{Instituto de Astrof\'{\i}sica de Canarias}

\author[0000-0002-7521-1078]{Kamoliddin Ergashev}
\affiliation{Ulugh Beg Astronomical Institute, Tashkent, Uzbekistan}

\author[0000-0002-4767-9861]{Tony Farnham}
\affiliation{University of Maryland}

\author{Petr Fatka}
\affiliation{Astronomical Institute of the Academy of Sciences of the Czech Republic, Fri\v{c}ova 298, Ond\v{r}ejov, CZ-25165, Czech Republic}

\author[0000-0002-0535-652X]{Marin Ferrais} 
\affiliation{Florida Space Institute, University of Central Florida, 12354 Research Parkway, Orlando, FL, 32826, USA}

\author[0000-0001-9456-8358]{Stefan Geier}
\affiliation{Gran Telescopio Canarias (GRANTECAN), Cuesta de San Jos\'e s/n, E-38712, Bre\~na Baja, La Palma, Spain}
\affiliation{Instituto de Astrof\'isica de Canarias, V\'ia L\'actea s/n, E38200, La Laguna, Tenerife, Spain}

\author[0000-0001-5749-1507]{Edward Gomez}
\affiliation{Las Cumbres Observatory}
\affiliation{School of Physics and Astronomy, Cardiff University, Queens Buildings, The Parade, Cardiff CF24 3AA, UK.}

\author[0000-0002-4439-1539]{Sarah Greenstreet}
\affiliation{DiRAC Institute and the Department of Astronomy, University of Washington}

\author{Hannes Gr\"{o}ller}
\affiliation{LPL/UA}

\author{Carl Hergenrother}
\affiliation{Ascending Node Technologies, LLC, Tucson, Arizona, USA }

\author[0000-0002-4043-6445]{Carrie Holt}
\affiliation{University of Maryland}

\author{Kamil Hornoch}
\affiliation{Astronomical Institute of the Academy of Sciences of the Czech Republic, Fri\v{c}ova 298, Ond\v{r}ejov, CZ-25165, Czech Republic}

\author[0000-0002-5932-7214]{Marek Hus\'{a}rik}
\affiliation{Astronomical Institute of the Slovak Academy of Sciences, 059 60 Tatransk\'{a} Lomnica, The Slovak Republic}

\author[0000-0002-6653-0915]{Raguli Inasaridze}
\affiliation{E. Kharadze Georgian National Astrophysical Observatory, Abastumani, Georgia}
\affiliation{Samtskhe-Javakheti State University, Akhaltsikhe, Georgia}

\author[0000-0001-8923-488X]{Emmanuel Jehin}
\affiliation{Space sciences, Technologies \& Astrophysics Research (STAR) Institute, University of Li\`{e}ge, Belgium}

\author[0000-0001-5098-4165]{Elahe Khalouei} 
\affiliation{Astronomy Research Center, Research Institute of Basic Sciences, Seoul National University, 1 Gwanak-ro, Gwanak-gu, Seoul 08826, Korea}

\author{Jean-Baptiste Kikwaya Eluo}
\affiliation{Vatican Observatory, V-00120 Vatican City of State}

\author[0000-0002-4787-6769]{Myung-Jin Kim}
\affiliation{Korea Astronomy and Space Science Institute, 776, Daedeokdae-ro, Yuseong-gu, Daejeon 34055, Korea}

\author[0000-0002-3171-9873]{Yurij Krugly}
\affiliation{Institute of Astronomy, V.N. Karazin Kharkiv National University, Kharkiv, Ukraine}
\affiliation{Astronomical Observatory Institute, Faculty of Physics, Adam Mickiewicz University, S{\l}oneczna 36, 60-286 Pozna{\'n}, Poland}

\author{Hana Ku\v{c}\'akov\'a}
\affiliation{Astronomical Institute of the Academy of Sciences of the Czech Republic, Fri\v{c}ova 298, Ond\v{r}ejov, CZ-25165, Czech Republic}

\author{Peter Ku\v{s}nir\'ak}
\affiliation{Astronomical Institute of the Academy of Sciences of the Czech Republic, Fri\v{c}ova 298, Ond\v{r}ejov, CZ-25165, Czech Republic}

\author[0000-0002-0772-0225]{Jeffrey A. Larsen}
\affiliation{USNA}

\author[0000-0002-6839-075X]{Hee-Jae Lee}
\affiliation{Korea Astronomy and Space Science Institute, 776, Daedeokdae-ro, Yuseong-gu, Daejeon 34055, Korea}

\author[0000-0003-0165-7701]{Cassandra Lejoly}
\affiliation{LPL/UA}

\author{Javier Licandro}
\affiliation{Instituto de Astrof\'{\i}sica de Canarias}

\author[0000-0001-9330-5003]{Pen{\'e}lope Longa-Pe{\~n}a}
\affiliation{Centro de Astronom{\'{\i}}a, Universidad de Antofagasta, Av.\ Angamos 601, Antofagasta, Chile}

\author{Ronald A. Mastaler}
\affiliation{LPL/UA}

\author[0000-0001-5807-7893]{Curtis McCully}
\affiliation{Las Cumbres Observatory}

\author[0000-0001-5666-9967]{Hong-Kyu Moon}
\affiliation{Korea Astronomy and Space Science Institute, 776, Daedeokdae-ro, Yuseong-gu, Daejeon 34055, Korea}

\author[0000-0003-2535-3091]{Nidia Morrell}
\affiliation{Las Campanas Observatory, Chile}

\author[0009-0007-5329-9148]{Arushi Nath}
\affiliation{MonitorMyPlanet, Toronto, Canada}

\author[0000-0002-5356-6433]{Dagmara Oszkiewicz}
\affiliation{Astronomical Observatory Institute, Faculty of Physics, Adam Mickiewicz University, S{\l}oneczna 36, 60-286 Pozna{\'n}, Poland}

\author{Daniel Parrott} 
\affiliation{Tycho Tracker}

\author{Liz Phillips}
\affiliation{Las Cumbres Observatory}
\affiliation{Department of Physics, University of California, Santa Barbara}

\author{Marcel M. Popescu}
\affiliation{Astronomical Institute of the Romanian Academy, 5 Cu\c{t}itul de Argint, 040557 Bucharest, Romania}

\author{Donald Pray}
\affiliation{Sugarloaf Mountain Observatory, South Deerfield, MA, USA}

\author{George Pantelimon Prodan}
\affiliation{Astronomical Institute of the Romanian Academy, 5 Cu\c{t}itul de Argint, 040557 Bucharest, Romania}

\author[0000-0003-2935-7196]{Markus Rabus} 
\affiliation{Departamento de Matem\'atica y F\'isica Aplicadas, Facultad de Ingenier\'ia,  Universidad Cat\'olica de la Sant\'isima Concepci\'on, Alonso de Rivera 2850, Concepci\'on, Chile}

\author{Michael T. Read}
\affiliation{LPL/UA}

\author[0000-0001-9944-8398]{Inna Reva}
\affiliation{Fesenkov Astrophysical Institute, Almaty, Kazakhstan}

\author{Vernon Roark}
\affiliation{University of Hawaii}

\author{Toni Santana-Ros}
\affiliation{Departamento de Física, Ingeniería de Sistemas y Teoría de la Señal, Universidad de Alicante, Carr. de San Vicente del Raspeig, s/n, 03690 San Vicente del Raspeig, Alicante, Spain}
\affiliation{Institut de Ciències del Cosmos (ICCUB), Universitat de Barcelona (IEEC-UB), Carrer de Martí i Franquès, 1, 08028, Barcelona, Spain}

\author{James V. Scotti}
\affiliation{LPL/UA}

\author{Taiyo Tatara}
\affiliation{USNA}

\author[0000-0002-1506-4248]{Audrey Thirouin}
\affiliation{Lowell Observatory, 1400 West Mars Hill Road, Flagstaff, AZ 86004, USA}

\author{David Tholen}
\affiliation{University of Hawaii}

\author[0000-0002-5899-2300]{Volodymyr Troianskyi}
\affiliation{Astronomical Observatory Institute, Faculty of Physics, Adam Mickiewicz University, S{\l}oneczna 36, 60-286 Pozna{\'n}, Poland}
\affiliation{Department of Physics and Astronomy FMPIT of Odesa I. I. Mechnykov National University, Pastera Street 42, 65082 Odesa, Ukraine}
\affiliation{Department of Physics and Methods of Teaching, Faculty of Physics and Technology, Vasyl Stefanyk Precarpathian National University, Shevchenko Street 57, 76000 Ivano-Frankivsk, Ukraine}

\author{Andrew F. Tubbiolo}
\affiliation{LPL/UA}

\author{Katelyn Villa}
\affiliation{Physics Department, United States Naval Academy, 572C Holloway Rd, Annapolis, MD 21402, USA}



\begin{abstract}

On 26 September 2022, the Double Asteroid Redirection Test (DART) spacecraft impacted Dimorphos, the satellite of binary near-Earth asteroid (65803) Didymos. This demonstrated the efficacy of a kinetic impactor for planetary defense by changing the orbital period of Dimorphos by 33 minutes \citep{Thomas23}. Measuring the period change relied heavily on a coordinated campaign of lightcurve photometry designed to detect mutual events (occultations and eclipses) as a direct probe of the satellite's orbital period. A total of 28 telescopes contributed 224 individual lightcurves during the impact apparition from July 2022 to February 2023. We focus here on decomposable lightcurves, i.e. those from which mutual events could be extracted.  We describe our process of lightcurve decomposition and use that to release the full data set for future analysis. We leverage these data to place constraints on the post-impact evolution of ejecta. The measured depths of mutual events relative to models showed that the ejecta became optically thin within the first $\sim1$ day after impact, and then faded with a decay time of about 25 days. The bulk magnitude of the system showed that ejecta no longer contributed measurable brightness enhancement after about 20 days post-impact. This bulk photometric behavior was not well represented by an HG photometric model. An HG$_1$G$_2$ model did fit the data well across a wide range of phase angles. Lastly, we note the presence of an ejecta tail through at least March 2023. Its persistence implied ongoing escape of ejecta from the system many months after DART impact.

\end{abstract}



%%%%%%%%%%%%%%%%
% INTRO
%%%%%%%%%%%%%%%%
\section{Introduction} \label{sec:intro}

Binary systems are estimated to represent about 15\% of the near-Earth asteroid population \citep{Pravec06}. Discovered as a binary in November 2003 \citep{Pravec03}, the near-Earth asteroid (65803) Didymos is an $\sim$760 m oblate spheroid with a $\sim150$ m satellite known as Dimorphos \citep{Naidu20,Daly23}. Based on extensive lightcurve \citep{Pravec22} and radar \citep{Naidu20} observations, the binary dynamics of this system have been well established \citep{Naidu22,Scheirich22}. Didymos has a rotation period = 2.2600 $\pm$ 0.0001 h and Dimorphos had an orbit period = 11.921481 $\pm$ 0.000016 h \citep{Naidu22}. This orbit period uncertainty of $<60$ ms makes Didymos one of the best characterized binary asteroids in the Solar System. Such precision was achievable because Didymos is an eclipsing binary. Mutual events -- occultations and eclipses -- can be detected in time series photometry of Didymos, and thus can serve as a chronometer for the orbital period of Dimorphos.

Given the state of knowledge of the Didymos system and favorable observing apparitions in the 2020's, this system was selected as the target for NASA's DART (Double Asteroid Redirection Test) mission \citep{Cheng16,Rivkin21}. Following its launch from Vandenberg Space Force Base on 24 November 2021 and a relatively short 10 month cruise phase, the DART spacecraft intentionally impacted Dimorphos on 26 September 2022 (at JD 2459849.46834). This was the world’s first full-scale planetary defense experiment and was designed to change the orbital period of Dimorphos as a test of asteroid deflection via kinetic impactor. In terms of Level 1 mission requirements \citep{Rivkin21}, the impact by DART was to change the orbit period by at least 73 s, which would then be measured via ground-based observations to a precision of 10\% or 7.3 s (0.002 h). DART impacted Dimorphos head-on \citep{Daly23} so that its orbital period decreased.

The DART spacecraft had a relatively simple payload that included a high-resolution imager called DRACO \citep[the Didymos Reconnaissance and Asteroid Camera for Optical navigation,][]{Fletcher18} and a 6U CubeSat called LICIACube \citep[the Light Italian CubeSat for Imaging of Asteroids,][]{Dotto21}, built by the Italian Space Agency (ASI). LICIACube separated from the DART spacecraft two weeks before impact and provided flyby imagery of the impact ejecta plume from about 30 s to 320 s after impact \citep{Dotto23}. Following these in situ operations, continued characterization of the system relied on remote telescopic observations.

An extensive campaign of ground and space-based observations was coordinated to study the aftermath of the DART impact and to meet the Level 1 requirements of the mission. The primary component of this campaign involved lightcurve photometry and the measurement of mutual events (\S\ref{sec:obs}). Based on the analysis of lightcurves from prior apparitions \citep{Pravec22}, the methodology for this campaign was well established. In short, high quality photometry (rms residuals generally $<0.01$ mag) was needed to enable the decomposition of lightcurves (\S\ref{sec:analysis}) into their constituent parts: the 2.26 h rotation of Didymos, a possible rotational signature from Dimorphos, drops in flux due to mutual events, and, in the post-impact environment, the evolution of ejecta. These stringent data requirements had to be sustained across many facilities, across many hours for each lightcurve, and across the duration of the apparition. Observing circumstances such as apparent magnitude and declination influenced  campaign planning. For example, large aperture facilities were primarily used at the beginning and end of the apparition when Didymos was faintest. Overall, this coordinated approach proved highly successful, yielding from just the first month of post-impact data a new orbital period for Dimorphos = 11.372 $\pm$ 0.017 h, corresponding to a change of -33 minutes relative to the pre-impact value \citep{Thomas23}. In this work we expand the scope of the lightcurve data set from that presented in \citet{Thomas23} to now include a full 8 months of data across the 2022-2023 apparition.

For definitional purposes we hereafter refer to a {\it lightcurve} as a time series of photometry collected by a single facility on a single night. A {\it lightcurve session} or {\it observing run} refers to the window in which a single lightcurve was obtained. We refer to the {\it primary lightcurve} as the rotational signature of Didymos. The term {\it secondary} refers to Dimorphos. Mutual events come in four flavors: {\it secondary eclipses} (Dimorphos passes into shadow), {\it secondary occultations} (Dimorphos moves behind Didymos), {\it primary eclipses} (the shadow of Dimorphos passes over Didymos), and {\it primary occultations} (Didymos is covered by Dimorphos).

In total, 28 observatories contributed data that were accepted as part of the DART lightcurve campaign (\S\ref{sec:facilities}). This produced a massive data set of 224 lightcurves with hundreds of mutual events detected from July 2022 to February of 2023. The associated decompositions (\S\ref{sec:results}) provided a basis for detailed modeling of the orbital and rotational dynamics in the Didymos system \citep[][]{Naidu23,Scheirich23}. We note that these two modeling efforts represent independent assessments of the lightcurve data set. The lightcurves presented here are a super set of the data analyzed in \citet{Scheirich23}, because their data quality requirements were more stringent, resulting in 193 lightcurves accepted for their analysis. These two analyses were meant to be completely independent, so it is expected that different acceptance criteria were applied. However, the \citet{Naidu23} and \citet{Scheirich23} orbit solutions agree within formal uncertainties, a good indication that the less stringent approach here did not bias results.

Our primary objective here is to provide an overview of the lightcurve campaign, associated data sets, and analysis, however we also leverage these data to address the evolution of post-impact ejecta. The measured depths of mutual events served as a proxy for the optical depth and fading of ejecta (\S\ref{sec:depth}). The photometry of Didymos, averaged over lightcurve variations, allowed for characterizing the photometric phase curve and determining when ejecta no longer contributed significant flux to the system (\S\ref{sec:mags}). Lastly, ejecta in the form of an extend tail persisted through to the end of the apparition; we quantify the tail's contribution to the total flux as a function of time (\S\ref{sec:tail}). Modeling of the dynamics of the Didymos system and refinements to the orbital period change of Dimorphos are presented elsewhere \citep[e.g.][]{Naidu23,Scheirich23}. Discussion of our results and prospects for future work serve as a conclusion to this paper (\S\ref{sec:disc}).

%%%%%%%%%%%%%%%%
% CAMPAIGN
%%%%%%%%%%%%%%%%
\section{DART Lightcurve Campaign} \label{sec:obs}

\begin{figure}[h!]
    \centering
    \includegraphics[width=\textwidth]{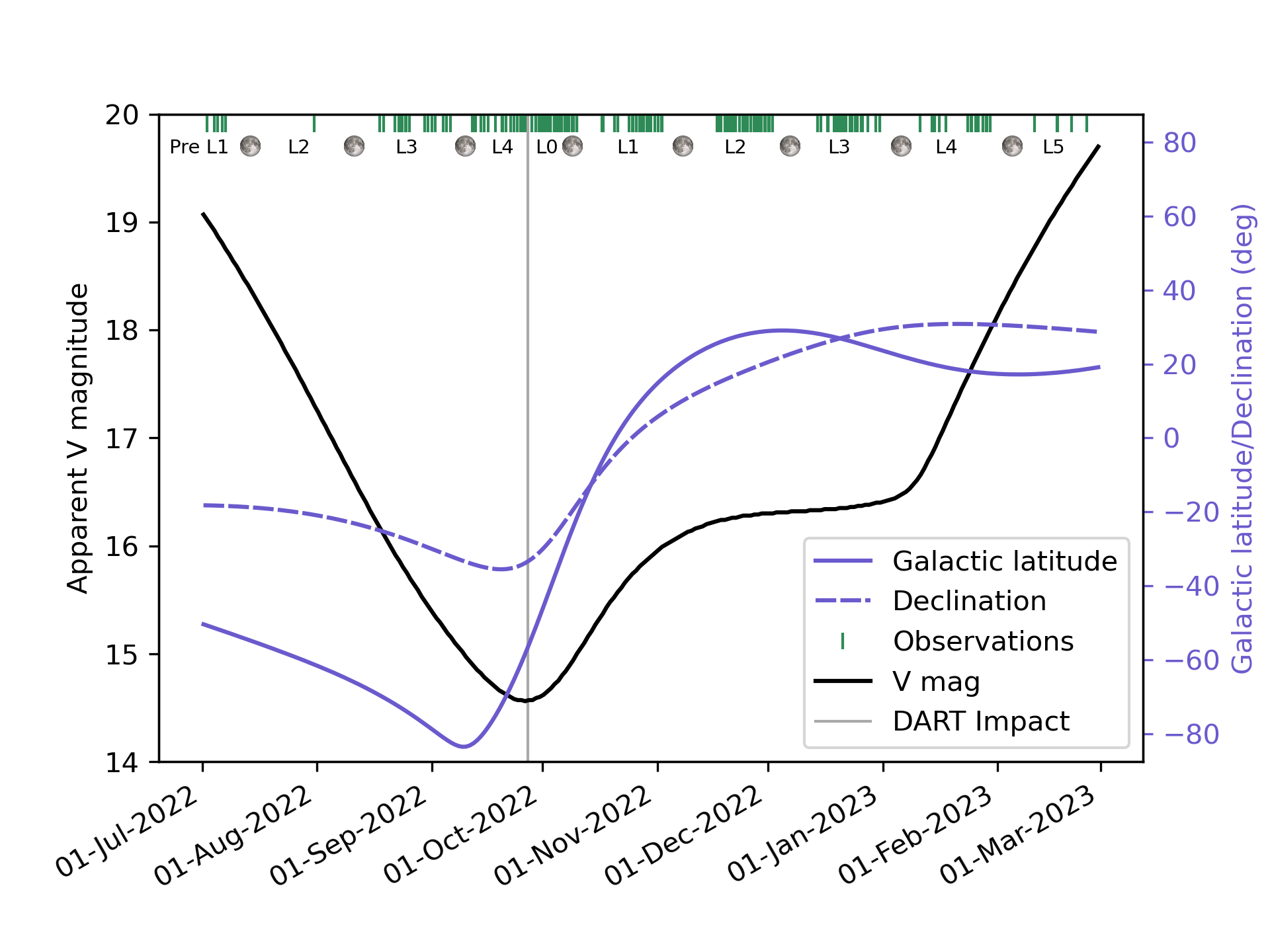}
    \caption{Observing circumstances for Didymos in the 2022-2023 apparition. The start dates of individual lightcurves (Table \ref{tab:details}) are indicated across the top as green vertical bars. Labels for pre and post impact lunations are shown along with full moon dates. The apparent magnitude, galactic latitude, and declination of Didymos were calculated from Lowell Observatory's \emph{astorb} system \citep{Moskovitz23}. Didymos reached a minimum magnitude of $V=14.5$ coincident with the DART impact on 26 September 2022 (vertical gray line). Gaps in lightcurve coverage were generally due to the full moon, however a galactic plane crossing in mid October also affected the number of viable lightcurves.}
    \label{fig:obs}
\end{figure}

The Didymos system underwent a highly favorable apparition in 2022-2023 (Figure~\ref{fig:obs}). This apparition was unusually long, with Didymos positioned at solar elongations greater than 100$^\circ$ for nearly 11 months from May 2022 to April 2023. During this time the system was predicted to reach a peak brightness of $V=14.5\,\rm{mag}$ at the end of September 2022, coincident with the DART impact. Didymos only gets this bright every few decades. The last time it was brighter than 15th magnitude was in 2003 when Dimorphos was discovered \citep{Pravec03}. It will not get this bright again until October 2062. At these magnitudes, obtaining high quality photometry is possible across a wide range of telescope apertures. We show in the following sections that telescope apertures down to 0.5 m in diameter were able to achieve strict data quality requirements and thus made significant contributions to the lightcurve campaign. Achieving the mission's level 1 requirement of measuring the orbital period change of Dimorphos from ground-based facilities \citep{Rivkin21} was largely possible because of the system brightness in this  apparition. 

The viewing geometry of Didymos in the impact apparition was in some ways advantageous. For example, a wide range of solar phase angles, from a maximum of 76$^\circ$ to a minimum of 6$^\circ$, enabled photometric (e.g. \S\ref{sec:mags}) and polarimetric \citep[e.g.][]{Bagnulo23} phase curves analyses. However, the ranges of declination and galactic latitude (Figure \ref{fig:obs}) posed interesting challenges. Given moderate negative declinations (around $-35^\circ$) in the days after impact followed by a transition into northern declinations in late October, the observing campaign necessitated a global approach that leveraged telescopes in both the northern and southern hemispheres. A galactic plane crossing 24 days after impact (on 20 October 2022) was expected to degrade the quality of photometry due to contamination by background sources for at least a week or two in late October. This turned out not to be a significant issue as viable lightcurves were obtained throughout the galactic plane crossing (\S\ref{sec:analysis}). More problematic was the high background and low elongation from a full moon on 9 October 2022. This led to a gap in viable lightcurves for about a week in the middle of the month. Fortunately, the post-impact brightening of Didymos by about $1.5\,\rm{mag}$ \citep{Graykowski23} allowed for short exposure times and thus helped to mitigate these issues of crowded fields and high background.

Taking these observational factors into account, the mission developed a 2022-2023 observing plan that spanned 9 individual lunations, with the impact lunation split into pre and post DART impact windows (Table \ref{tab:lunations}). The lunations represented windows outside of full moon conditions when the highest quality data was likely to be obtained. These covered the full apparition starting in July 2022 with pre-impact lunations L1 through L4, and then post-impact lunations that increased from L0 to L5, ending in February 2023. Data were obtained after the L5 lunation in March of 2023, but insufficient temporal coverage and low signal-to-noise rendered these data unusable for our analysis here. Unsurprisingly, the number of viable lightcurves and associated decompositions (Table \ref{tab:lunations}) tracked inversely with the apparent magnitude of Didymos. At the beginning (Pre L1) and end (L5) of the apparition when Didymos was faintest, telescopes with apertures $>$2m in diameter were required to collect data of sufficient quality. The primary goal of observations in the pre-impact lunations was to confirm the dynamics of the system as determined by \citet{Naidu22} and \citet{Scheirich22}. There was also interest in measuring the rotation period of Dimorphos in the pre-impact data, but that signature was never clearly detected, perhaps due to its oblate shape \citep{Daly23}.

\begin{deluxetable*}{c  l  l  c  c}
\tablecaption{DART campaign lunations. The range of dates in which data were obtained are given for pre-impact lunations L1-L4 and post-impact lunations L0-L5. The number of individual lightcurves and decompositions are given for each lunation. \label{tab:lunations}}
\tabletypesize{\footnotesize}
\tablehead{\colhead{Lunation} & \colhead{Data Range (JD)} & \colhead{Data Range (UTC)} & \colhead{Lightcurves} & \colhead{Decompositions}}
\startdata
Pre L1  &   2459762.6669 - 2459767.9728   & 2022-07-02T04:00 - 2022-07-07T11:20 & 5 & 1\\
Pre L2  &   2459791.6504 - 2459791.9057   & 2022-07-31T03:36 - 2022-07-31T09:44 & 1 & 1 \\
Pre L3  &   2459809.5313 - 2459828.6771   & 2022-08-18T00:45 - 2022-09-06T04:15 & 16 & 4 \\
Pre L4  &   2459834.5636 - 2459848.8959   & 2022-09-12T01:31 - 2022-09-26T09:30 & 22 & 3 \\
L0      &   2459850.6063 - 2459862.7933   & 2022-09-28T02:33 - 2022-10-10T07:02 & 54 & 13 \\
L1      &   2459869.6181 - 2459886.0129   & 2022-10-17T02:50 - 2022-11-02T12:18 & 27 & 4 \\
L2      &   2459900.7059 - 2459915.9442   & 2022-11-17T04:56 - 2022-12-02T10:39 & 48 & 7 \\
L3      &   2459927.9180 - 2459944.8952   & 2022-12-14T10:01 - 2022-12-31T09:29 & 30 & 6 \\
L4      &   2459955.5764 - 2459974.7416   & 2023-01-11T01:50 - 2023-01-30T05:47 & 16 & 2 \\
L5      &   2459986.5711 - 2460000.8285   & 2023-02-11T01:42 - 2023-02-25T07:53 & 5 & 2 \\
\enddata
\end{deluxetable*}

For all lunations the mission established strict data quality requirements to ensure successful lightcurve decompositions (\S\ref{sec:analysis}). These requirements were largely based on experience gained from previous apparitions in 2003, 2015, 2017, 2019, 2020, and 2021 \citep[e.g.][]{Pravec22,Naidu22,Scheirich22}. Requirements provided to observers leading up to the apparition included photometric precision of at least 0.01 mag per exposure, and a temporal cadence of no more than 180 s between exposures to ensure adequate sampling of mutual events. The choice of photometric filter was not prescribed, because our primary photometric analysis was based on differential magnitudes and we assumed that Didymos had no rotational color variability that would affect combining data from different filters. Observers were encouraged to employ whatever filter would yield the highest signal-to-noise with their instrument. Typically, lightcurves were accepted for analysis only if they spanned at least one full rotation of Didymos (2.26 h), though exceptions were made for some short lightcurve segments when data from other facilities were taken close enough in time to enable a clean decomposition. These short segments were often useful in filling out specific rotational phases of Didymos that were not covered by adjacent lightcurves. Decompositions were performed on batches of data in which the morphology of the Didymos lightcurve was roughly constant. We discuss this process in greater detail in Section \ref{sec:analysis}.

Though observations were planned in the days immediately after impact to monitor the evolution of the ejecta plume, the clearing of ejecta and its contribution to the measured photometry of the system was expected to confuse the detection of mutual events for many days or even weeks after impact \citep{Fahnestock22}. Given the full moon on October 9 and the galactic plane crossing a little over a week later (Figure \ref{fig:obs}), it was unclear whether mutual event detections and thus a period change determination would be possible in the first month after impact. Thus the lightcurve campaign had to be both comprehensive (e.g. facilities spanning a range of apertures and locations on Earth) and flexible (e.g. in telescope scheduling) to effectively respond to whatever was the outcome of the impact experiment.

To help observers prepare for post-impact observational challenges, such as low galactic latitude and a bright moon, a list of practice targets was provided. This list was intended to help test and refine the capabilities of instruments under challenging conditions, and to help establish data analysis procedures that would enable rapid turn-around of reduced lightcurves. Test targets were selected from the catalog of known asteroids in the \emph{astorb} database \citep{Moskovitz23} that had analogous observing conditions to Didymos in October 2022. Specifically, we selected near-Earth objects with $ 14.5 < V < 16.5$, solar elongation $>90^\circ$, and non-sidereal rates of motion between 1.25 and 7.5 \arcsec/min. This list of test targets was posted online and dynamically updated on a daily basis to incorporate newly discovered objects and to handle changes in observability. This list was additionally sub-divided based on conditions related to lunar phase, lunar elongation, and galactic latitude. Three windows of observing conditions were defined: (1) lunar phase $>75\%$, lunar elongation $>60^\circ$, and galactic latitude $>20^\circ$, (2) $25\%<$ lunar phase $<75\%$, $30^\circ <$ lunar elongation $<50^\circ$, and galactic latitude $<10^\circ$, and (3) lunar phase $<25\%$, lunar elongation $>45^\circ$, and galactic latitude $<20^\circ$. These three windows represented the conditions for Didymos from October 1-11, October 15-19, and October 20-27, respectively. For the year leading up to impact any given night was likely to have $\sim1-3$ test targets available that met these conditions.

The full extent of the 2022-2023 campaign involved contributions from many observers and facilities across the globe. Details of the 224 individual lightcurves presented in the remainder of this work are given in Appendix Table \ref{tab:details}. For completeness we include here data collected by the  investigation team from July 2022 to February 2023. Some of these observations (UT 2-7 July 2022, 28 September - 10 October 2022) were previously reported in \citet{Thomas23}. We present only those data that met the quality requirements described above. Many dozens of additional data sets (about 25\% of those submitted by the investigation team) were unfortunately not accepted as viable for lightcurve analysis. However, some of these data sets have and will continue to provide valuable insights into other aspects of the post-impact Didymos environment \citep[e.g.][]{Kareta23}.

%%%%%%%%%%%%%%%%
% TELESCOPES
%%%%%%%%%%%%%%%%

\section{Telescope Facilities\label{sec:facilities}}

\begin{figure}[h!]
    \centering
    \includegraphics[width=\textwidth]{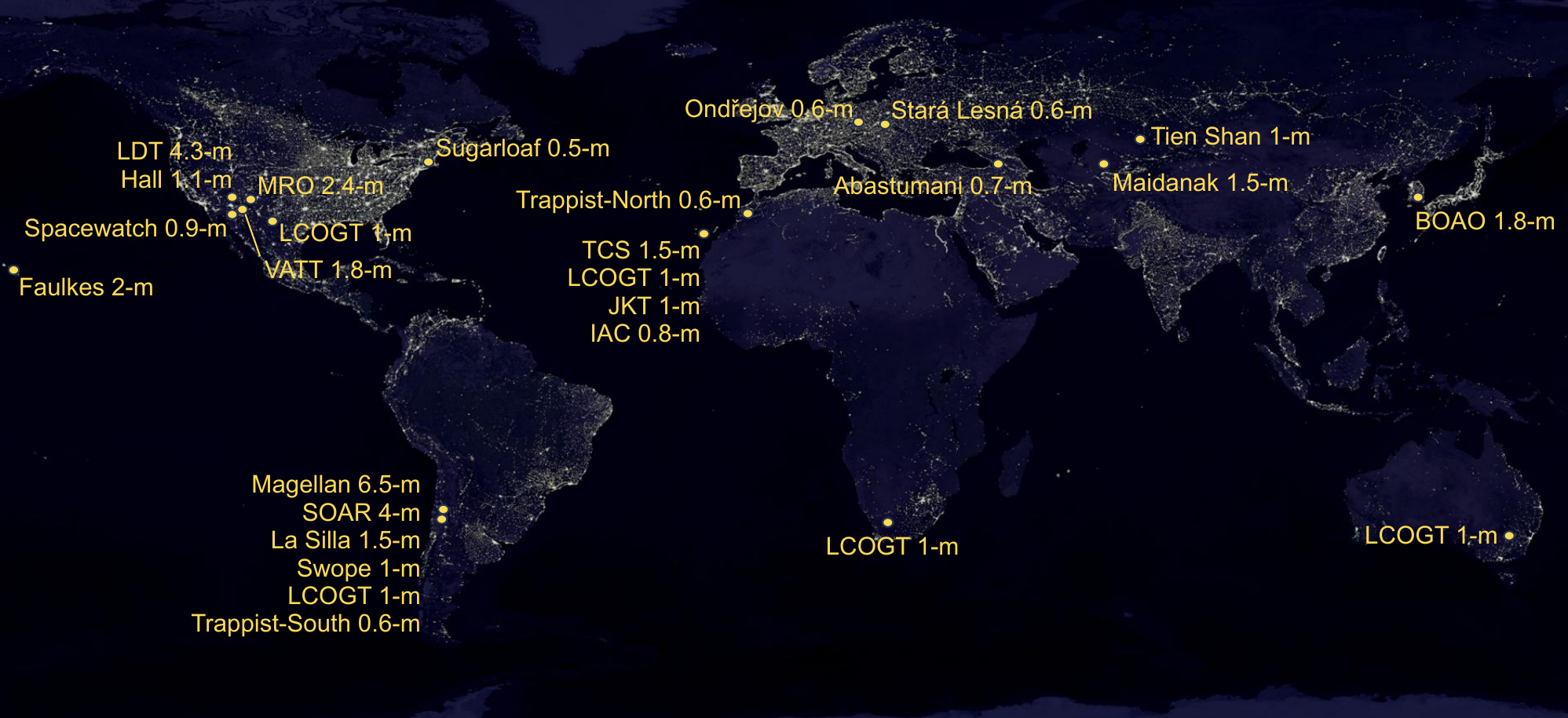}
    \caption{Global distribution of telescopes that contributed lightcurves to the 2022-2023 campaign. See text for details of facilities and instruments. Earth at night image credit: NASA/NOAA.}
    \label{fig:map}
\end{figure}

A total of 28 telescopes contributed viable data to the lightcurve campaign (Figure \ref{fig:map}, Table \ref{tab:observatories}). These telescopes ranged in diameter from 0.5 m up to 6.5 m and employed a wide variety of instruments. In addition, the photometric filters and tracking modes (sidereal vs non-sidereal) varied from one facility to the next. This approach to building the lightcurve data set was a natural consequence of the diversity and scope of the DART investigation team, and may have helped to minimize systematic biases that could have affected outcomes if fewer facilities were involved. Careful control of systematics and data quality were essential to the success of the campaign, and were the primary challenge to building the full data set. Though the general methodology of collecting images with CCD cameras and measuring lightcurves is hardly novel, the DART campaign required that this be done at high precision (sub-percent photometry), across a large number of facilities, and be sustained for multiple hours within a night, and across many months throughout the apparition. 

To achieve these high standards, individual observers were encouraged to adopt whatever reduction methods worked best for their data. However, some aspects of the reductions were common to most data sets. In the post-impact window, aperture sizes significantly larger than the local seeing (5--7\arcsec~radius for many of the data sets in October) were typically used to account for the extended brightness of the ejecta cloud. This was important to compensate for centroiding errors related to the complicated point spread function of the post-impact system. It was found that the larger apertures generally caused higher noise levels in individual data points, but much better point-to-point consistency across each lightcurve. In nearly all cases circular aperture photometry was employed. Reductions involved testing a range of photometric aperture sizes to optimize both the signal-to-noise of individual measurements as well as the consistency of intra-night measurements. In general, the magnitudes reported by each facility were converted to differential values by subtracting off the mean of each lightcurve outside of mutual events. Though error bars were reported for most lightcurves, these were not measured in a consistent way across all data sets, and thus were largely ignored in our analysis.

 Some telescopes elected to track at half of the non-sidereal rates of the asteroid. This was particularly true around the time of minimum geocentric distance when the non-sidereal rates reached a maximum of $\sim8$\arcsec/min. Fortunately the asteroid was also brightest at this time, thus long exposures were not required. Generally, exposures times were kept below the level where significant elongation of the point spread function occurred, eliminating the need for non-circular apertures. Noise characteristics for these data were dominated by the signal from the asteroid (and its morphologically complex ejecta cloud), as opposed to being background limited. Thus the use of circular apertures for these data, when the asteroid and stars may have been slightly trailed, did not introduce significant background noise into the measurements.

As a way to ensure that the period change measurement was adequately supported, the DART project contracted several facilities to carry out lightcurve observations. These included the 6.5-m Magellan Baade and 1-m Swope telescopes at Las Campanas Observatory in Chile, the 2.4-m Magdalena Ridge Observatory in New Mexico, the 4.3-m Lowell Discovery Telescope in Arizona, and the Las Cumbres Observatory Global Telescope network of 1-m facilities. All other facilities that contributed to the lightcurve campaign were not directly contracted by the project. These unsupported observatories contributed a majority of the lightcurves to the overall data set and serve as a testament to the global interest in the DART experiment.

In the following sub-sections we summarize in order of aperture size the telescope, instrument, and reduction methods used by each facility. Details on individual lightcurves are in Appendix Table \ref{tab:details}. In addition, a large file is included with this manuscript as supporting data that contains all of the individual lightcurve measurements (38532 in total!) and the associated decomposed residuals that were used for mutual event analysis (\S\ref{sec:analysis}). All original fits files from the contracted facilities will be made publicly available through NASA's Planetary Data System (PDS) Small Body Node.

\begin{table}[htbp]
\centering
\footnotesize
\rotatebox{90}{
\begin{tabular}{ llllr } 
 \hline
Telescope & Instrument & Location & IAU Code(s) & \# lightcurves \\
  \hline
6.5-m Magellan Baade & IMACS & Las Campanas, Chile & 269 & 1 \\
4.3-m LDT & LMI & Happy Jack, Arizona, USA & G37 & 8 \\
4.1-m SOAR & Goodman & Cerro Pach\'{o}n, Chile & I33 & 2 \\
2.4-m MRO & MRO2k CCD & Magdalena Ridge, New Mexico, USA & H01 & 11 \\
2.0-m Faulkes North & MuSCAT3 & Haleakala, Hawaii, USA & F65 & 1 \\
1.8-m VATT & STA 4k CCD & Mount Graham, Arizona, USA & 290 & 2 \\
1.8-m BOAO & e2v 4k CCD & Bohyunsan, South Korea & 344 & 2\\
1.5-m Danish Telescope & DFOSC & La Silla, Chile & W74 & 42\\
1.5-m AZT-22 Telescope & SNUCAM & Maidanak Observatory, Uzbekistan & 188 & 2 \\
1.5-m TCS & MuSCAT2 & Tenerife, Spain & 954 & 4 \\
1.1-m Hall Telescope & NASA42 & Anderson Mesa, Arizona, USA & 688 & 27 \\
1-m LCOGT & Sinistro & McDonald Observatory, Texas, USA & V37, V39 & 5 \\
1-m LCOGT & Sinistro & Siding Spring, Australia & Q63, G64 & 2 \\
1-m LCOGT & Sinistro & Sutherland, South Africa & K91, K92, K93 & 9 \\
1-m LCOGT & Sinistro & Cerro Tololo, Chile & W85, W86, W87 & 11 \\
1-m LCOGT & Sinistro & Tenerife, Spain & Z31, Z24 & 10 \\
1-m JKT & Andor 2k CCD & La Palma, Spain & 950 & 2 \\ 
1-m Swope Telescope & e2v 4k CCD & Las Campanas, Chile & 304 & 19 \\
1-m Tien-Shan Telescope & Apogee 3k CCD & Tien-Shan, Kazakhstan & N42 & 1 \\
0.9-m Spacewatch & 4-CCD mosaic & Kitt Peak, Arizona, USA & 691 & 14 \\
0.8-m IAC80 & CAMELOT2 & Observatorio del Teide, Spain & 954 & 1 \\
0.7-m AC-32 Telescope & FLI 2k CCD & Abastumani , Georgia & 119 & 5 \\
0.6-m Ond\v{r}ejov Telescope & Moravian 2k CCD & Ond\v{r}ejov, Czech Republic & 557 & 4 \\
0.6-m Sugarloaf Telescope & SBIG 2k CCD & Deerfield, Massachusetts, USA & - & 4 \\
0.6-m G2 Telescope & FLI 2k CCD & Star\'{a} Lesn\'{a} Observatory, Slovakia & - & 1 \\
0.6-m TRAPPIST-North & Andor iKon-L BEX2 & Ouka\"{i}meden Observatory, Morocco & Z53 & 8 \\
0.6-m TRAPPIST-South & FLI ProLine 3041-BB & La Silla, Chile & I40 & 25 \\ 
0.5-m T72 iTelescope & FLI 4k CCD & Deep Sky Chile Observatory, Chile & X07 & 1 \\
  \hline
\end{tabular}
}
\caption{Details of facilities that contributed to the 2022-2023 lightcurve campaign. IAU assigned observatory codes are given when available. The number of individual lightcurves contributed by each facility are listed in the final column.}
\label{tab:observatories}
\end{table}

\subsection{6.5-m Magellan-Baade}

The Baade 6.5-m telescope is located at Las Campanas Observatory, on the Atacama desert in the north of Chile, at an elevation of 2400m. We employed the IMACS (Inamori-Magellan Areal Camera and Spectrograph) instrument \citep{Dressler11}, which is equipped with two arrays of eight 2k $\times$ 4k e2v detectors, where each array provides a different pixel scale. Only the IMACS-F2 array with detector \#2 was used for this project, which has 15 $\mu m$ pixels that each image 0.2 arcseconds when un-binned. The asteroid was observed with a Sloan-r filter and fixed pointings on the sky, letting the asteroid cross the detector's field while changing to a new pointing when necessary.

The IMACS-F2 raw images were processed in standard way, i.e. bias subtraction and flat fielding. Astrometry was performed using our own python scripts built using the {\it astropy} package. Aperture photometry was measured on the asteroid and a selection of the brightest stars in every pointing to estimate the individual image zero points and the differential photometry of the asteroid. The photometry was measured using our own python scripts and the SEP python library \citep{Barbary18} source extraction tools. For the zero points the stars were matched against the GAIA DR3 catalog \citep{Gaia21} when possible, and against the PanSTARRS catalog when not.

\subsection{4.3-m Lowell Discovery Telescope (LDT)}

The LDT is located in Happy Jack, Arizona at an elevation of 2360 m. All LDT images were obtained with the Large Monolithic Imager (LMI), a 6k $\times$ 6k e2v CCD with 15 $\mu m$ pixels. LMI images a 12\arcmin~field of view at an un-binned pixel scale of 0.12 \arcsec/pixel. All images were obtained in 3$\times$3 binning mode with a broad $VR$ filter that provided high throughput from about 500 to 700 nm. For all LDT observations the telescope was tracked at sidereal rates, allowing the asteroid to pass through fixed star fields. Multiple pointings were used in a single night when the motion of the asteroid exceeded the instrument field of view. Individual exposure times ranged from 15 to 160 seconds across the apparition.

The reduction of LMI images followed standard flat field and bias correction techniques. The photometry of Didymos was measured and calibrated using the Python-based Photometry Pipeline \citep[PP,][]{Mommert17} as described in \citet{Pravec22}. In summary, PP employed SourceExtractor \citep{Bertin96} to extract sources from the fields, Scamp \citep{Bertin06} to register the astrometry of those sources relative to the Gaia DR2 reference catalog \citep{Gaia18}, and then calibrated the photometry relative to the the PanSTARRS DR1 catalog \citep{Flewelling20}. Only field stars with solar-like colors (i.e. $g-r$ and $r-i$ colors within 0.2 magnitudes of the Sun) were used for photometric calibration. Typically more than 10 field stars were used to calibrate each image. A curve of growth analysis was performed each night to optimize the photometry aperture. This analysis aimed to optimize both the signal-to-noise of individual measurements as well as the consistency of intra-night measurements to minimize point-to-point scatter. Aperture radii ranged from 3.5 to 7 pixels (1.26\arcsec~to 2.52\arcsec) across the apparition.

\subsection{4.1-m Southern Astrophysical Research Telescope (SOAR)}

The 4.1-m SOAR telescope is located on Cerro Pachon in central Chile at an elevation of 2713 meters. Images were obtained with the Goodman spectrograph and imager \citep{Clemens04} which employs an e2v 231-84 CCD with 4k $\times$ 4k pixels. In imaging mode the CCD receives a 7.2\arcmin~circular field of view on a 3k $\times$ 3k portion of the chip. The unbinned pixel scale is 0.15 \arcsec/pixel. We operated the camera in 2$\times$2 binning mode with a $VR$ filter that provided high throughput from approximately 500 to 700 nm. Individual image exposure times were 90 seconds.

The reduction and measuring of photometry from the SOAR data followed an identical procedure to that used for LDT. The Photometry Pipeline referenced the Gaia DR2 \citep{Gaia18} and PanSTARRS \citep{Flewelling20} catalogs for astrometric and photometric calibration. Aperture radii of 7 pixels (2.1\arcsec) and 6 pixels (1.8\arcsec) were used on the nights of UT 4 July 2022 and 5 July 2022 respectively.

\subsection{2.4-m Magdalena Ridge Observatory (MRO)}

The Magdalena Ridge Observatory’s (MRO) fast-tracking 2.4-meter telescope is located at an elevation of 3250 meters in the Magdalena Mountains near Socorro, New Mexico. All MRO images were acquired with  “MRO2K”, which is an Andor iKon-L 936 camera operating at 188K, utilizing a 2048 $\times$ 2048 back-illuminated e2v CCD with 13.5 $\mu m$ pixels. The unbinned pixel scale is 0.13 \arcsec/pixel yielding a 4.5 arc-minute field of view. All lightcurve data were acquired in 4 $\times$ 4 binning mode using either the Bessell $R$ or broad band $VR$ filter while tracking on Didymos to maximize its signal. Observations early in the apparition required separate images of comparison star fields due to Didymos’ rapid non-sidereal motion. For this reason, this time period also necessitated having photometric sky conditions. Exposure times ranged from 15 to 150 seconds throughout the apparition.

MRO images were reduced according to standard dark, bias, and flat field correction techniques. The photometry of Didymos was measured using the IRAF {\it apphot} package \citep{1986SPIE..627..733T}. The instrumental magnitude of Didymos was measured in each field using apertures that ranged from 4-10 pixels (2.1-5.2 arc-seconds) depending on seeing conditions. In addition, an ensemble of typically 5-8 stars in each comparison field was also measured in either the same image or in a separate comparison star image. An initial analysis was performed on the comparison stars to assess their robustness. A temporally interpolated average magnitude of the comparison stars was then subtracted from each Didymos instrumental magnitude resulting in a differential magnitude. The resulting lightcurve magnitudes were then reported as “relative” with an arbitrary zero point.

\subsection{2.0-m Faulkes North}

Faulkes Telescope North (FTN) is located on Haleakala, Maui in Hawaii. FTN images were collected with MuSCAT3 \citep{Narita20}, a four channel simultaneous imager with $g$, $r$, $i$, and $z$ channels. The four independent channels employ 2k $\times$ 2k Princeton Instruments CCDs from the Pixis and Sophia model lines. Each CCD images a 9.1 arcminute field of view at a scale of 0.27 \arcsec/pixel. Individual exposure times were 30 seconds for all channels.

Reduction of the MuSCAT3 images employed the AstroImageJ \citep[AIJ,][]{Collins17} package. Within AIJ, the multiple aperture differential photometry tool was used to settle on an optimal aperture radius of 12 pixels, and a background annulus with an inner radius of 15 and an outer radius of 20 pixels. The measured fluxes from each of the 4 simultaneous $griz$ exposures were combined into a single arbitrary magnitude, and then calibrated against the Gaia DR3 catalog \citep{Gaia21}. Four field stars per frame with roughly solar-like colors were used for this calibration.

\subsection{1.8-m Vatican Advanced Technology Telescope (VATT)}

The Vatican Observatory's Vatican Advanced Technology Telescope (VATT) is located at Mount Graham, Arizona, and is an Aplanatic Gregorian 1.8-m f/9 telescope with a 0.38-m f/0.9 secondary mirror. The VATT4K CCD camera was used for all VATT observations and consists of a STA0500A 4096 $\times$ 4096-pixel back-illuminated detector with 15 $\times$ 15-micron pixels. VATT4K images have a 12.5\arcmin~field of view and were obtained in a 2$\times$2 binning mode yielding a binned pixel scale of 0.38\arcsec/pixel. 

For VATT data, the telescope was tracked at sidereal rates with multiple pointings in a single night to keep the asteroid within the field of view. Individual exposures of 30 and 60 seconds in duration were taken through a Harris $V$-band filter. The reduction of VATT4K images followed standard flat field and bias correction techniques. The observations were measured using the Tycho Tracker software \citep{Parrott20} with photometry calibrated relative to the ATLAS stellar catalog \citep{Tonry18}. Photometric calibration was derived from field stars with solar-like colors. A circular photometric aperture with a radius of 11 pixels (4.1\arcsec) was used in conjunction with a sky background annulus with an inner radius of 23 pixels (8.5\arcsec) and an outer radius of 33 pixels (12.2\arcsec). 

\subsection{1.8-m Bohyunsan Optical Astronomy Observatory (BOAO)}

The BOAO is located in Yeongcheon, Korea, at an altitude of 1143 m. Observations were conducted using the 1.8 m telescope at BOAO with an e2v 4k CCD and Cousins $R$-filter. All images were taken in 2 $\times$ 2 binning mode with an effective pixel scale of 0.43\arcsec/pixel, resulting in a field of view of 14.7\arcmin~$\times$ 14.7\arcmin. Individual exposure times were set to 100 seconds.

The images from BOAO were reduced using the Image Reduction and Analysis Facility (IRAF) software package. We performed calibration procedures, including bias, dark, and flat-field corrections, following standard protocols. To calculate the World Coordinate System (WCS) solution, we utilized the SCAMP package \citep{Bertin06} and matched fields to the Gaia DR2 catalog \citep{Gaia18}. Aperture photometry was conducted on these images using the IRAF/APPHOT package. The aperture radius was set to 10 pixels ($\sim4.5$\arcsec) to minimize point-to-point scatter. The photometric calibrations were performed following the method of \citet{Gilliland88}, namely using ensemble normalization employing standard magnitudes obtained from the Pan-STARRS Data Release 1 catalog \citep[PS DR1;][]{Flewelling20} . For consistency, we converted the PS DR1 magnitudes to the Johnson–Cousins filter magnitude system using empirical transformation equations \citep{Tonry12}.

\subsection{1.54-m Danish Telescope}

The 1.54-m Danish Telescope is located at La Silla Observatory, Chile at an elevation of 2366~m. It is operated jointly by the Niels Bohr Institute, University of Copenhagen, Denmark and the
Astronomical Institute of the Academy of Sciences of the Czech Republic. All images in the DART campaign were obtained by the Danish Faint Object Spectrograph and Camera (DFOSC) with an e2v CCD 231-41 sensor and standard Johnson-Cousins $V$ and $R$ photometric filters \citep{Bessell90}. The CCD sensor has $2048 \times 2048$ square pixels ($13.5~\mu$m size) and we used it in $1 \times 1$ binning mode to produce a scale of 0.396~arcsec/pixel and a $13.5 \times 13.5$~arcmin$^2$ field of view. For the observations taken in September and the first half of October 2022, the telescope was tracked at sidereal rates, allowing the asteroid to pass through fixed star fields.  For the observations taken from 2022-10-29 to 2023-01-29, the telescope was tracked at half the apparent rate of the asteroid, providing star and asteroid images of the same profile in one frame that facilitated obtaining robust photometric reduction.  Multiple pointings were used in a single night for the observations taken in September and the first half of October when the motion of the asteroid in a night exceeded the instrumental field of view. We used a single set of local reference stars for each night of observations from late October 2022 through late January 2023. Individual exposure times ranged from 6 to 150 seconds across the apparition.

The reduction of the images followed standard flat field and bias frame correction techniques. The photometry of Didymos was measured and calibrated using {\it Aphot}, a synthetic aperture photometry software developed by M.~Velen and P.~Pravec at Ond\v{r}ejov Observatory. It reduces asteroid images with respect to a set of field stars and the reference stars are then calibrated in the Johnson-Cousins photometric system using \citet{Landolt92} standard stars on a night with photometric sky conditions. This resulted in $R$ magnitude errors of about 0.01~mag.  Typically 8 local reference field stars, which were checked for stability (non-variable, not of extreme colors), were used on each night, or for each pointing on nights before mid-October.  Aperture radii from 6 to 10 pixels (2.4 to 4.0 arcsec) were found optimal on the individual nights.

\subsection{1.5-m AZT-22 telescope at Maidanak Observatory}

The 1.5 m AZT-22 telescope is located in the western part of Maidanak Mountain in the south of Uzbekistan at an elevation of 2593 m. The Didymos observations were carried out with the Seoul National University 4k $\times$ 4k CCD Camera (SNUCAM) which has 4096 $\times$ 4096 square 15 $\mu m$ pixels with a CCD chip manufactured by Fairchild Instruments \citep{Im10}. All images were obtained with an un-binned pixel scale 0.27 \arcsec/pixel and a field of view 18.1\arcmin~$\times$ 18.1\arcmin~through the $R$ filter. The telescope was tracked at sidereal rates and exposure times were set to 60 seconds. 

The primary reduction of images was performed in a standard way with master-bias and master-flats, the latter of which was constructed with twilight flats obtained on nearby night under photometric conditions. Aperture photometry was performed using MPO Canopus\footnote{\url{https://minplanobs.org/BdwPub/php/displayhome.php}}. The ATLAS catalog \citep{Tonry18} was used to obtain calibrated R magnitudes for the asteroid based on comparison stars with colors close to the Sun. Five solar type stars were used to calibrate the frames. The diameter of the aperture used for the comparison stars was 11 pixels or about 3\arcsec. The images of the asteroid were slightly trailed over the 60-second exposures, so asteroid measurements were made with an elliptical aperture of 11 $\times$ 13 pixels. These aperture dimensions were chosen to roughly approximate isophotes for the stars and asteroid.

\subsection{1.5-m Telescopio Carlos S\'{a}nchez (TCS)}

The TCS belongs to the Instituto de Astrof\'isica de Canarias (IAC) and is located at Teide Observatory (latitude: $28\degr\,18\arcmin\,01\farcs8$ N; longitude: $+16\degr\,30\arcmin\,39\farcs2$ W, and an altitude of 2387~m). Typical seeing for this location is in the range of 1.0 to $1.5 \arcsec$.  The observations were performed with the MuSCAT2 instrument \citep{2019JATIS...5a5001N} which is mounted on the Cassegrain focus of the telescope. A system of lenses reduces the focal length of the system to a ratio of $f/4.4$.

This instrument allows simultaneous photometric observations in four visible broad-band filters, namely \textit{g}\,(400\,-\,550), \textit{r}\,(550\,-\,700), \textit{i}\,(700\,-\,820) and \textit{z$_s$}\,(820\,-\,920)\,nm. At the end of each of the four channels there are independently controllable charge couple device (CCD) cameras (1024\,$\times$\,1024 pixels). They have a pixel size of $\sim$0.44\,arcsec$\cdot$ pixel$^{-1}$ and a field of view (FoV) of 7.4\,$\times$\,7.4\,arcmin$^2$. The telescope was tracked at sidereal rates with the asteroid crossing the entire field of view. Because of the small FoV, multiple pointings were needed during the same observing session. The single image exposure times were 15 sec for the first three sessions (UT 30 September 2022, 7 October 2022, 16 October 2022) and 30 sec for the last (UT 16 November 2022).

The pre-processing of the images included bias and flat-field corrections. The remaining background patterns were removed using the GNU Astro package \citep{gnuastro, noisechisel_segment_2019}. The lightcurves were first measured with PP \citep{Mommert17}. For astrometric registration we used the GAIA catalog \citep{Gaia18}.  We discarded all those images for which the astrometric registration failed (due to bad tracking or variable sky conditions). The PanSTARRS catalog \citep{Flewelling20} was used for photometric calibration. PP was run with a fixed aperture of 4.4 arcsec (10 pixels). 

A second data reduction was performed using the Image Reduction and Analysis Facility -- IRAF \citep{1986SPIE..627..733T} software. In order to improve the signal to noise ratio (SNR) we combined the images taken simultaneously by the four channels in a single one. Then, we performed differential photometry using the Aperture Photometry Package -- APPHOT from IRAF. The magnitudes were computed using an aperture of 4.4 arcsec, and nine comparison stars from the same FoV were used to compute the differential photometry. To further improve the SNR, we binned every four exposures into a single point. We then spliced lightcurve segments from different pointings into a single lightcurve. The offsets between the lightcurve segments were computed using the calibrated photometry derived with PP on the observations made with the \textit{g} filter.

\subsection{1.1-m Hall}

The 1.1-m Hall Telescope is located on Anderson Mesa, nine air miles southeast of Flagstaff, Arizona, at an elevation of 2203 m. All images were taken with the NASA42 camera, a custom built CCD camera with a 4K $\times$ 4K array of 15 $\mu m$ pixels. The image scale after applying 3 $\times$ 3 binning was 1.09 arcsec per pixel, with a field of view of 24\arcmin. Images were all taken through a broadband $VR$ filter. In the first month after the DART impact, the Didymos system was moving at a rate of over 6 arcseconds per minute. Therefore, tracking was done at half of the ephemeris rate of the asteroid, and typically three pointings were made during the night. Exposure times ranged from 90 to 180 seconds.
 
Data reduction began with image calibration with MaxIm DL\footnote{\url{https://diffractionlimited.com/product/maxim-dl/}}, using sets of 15 bias and flat frames that were typically collected at the beginning of each night. Groups of images at each pointing were astrometrically solved, registered, and aligned in MaxIm DL. Photometry was performed with MPO Canopus. Typically, five comparison stars with solar color ($B-V$ color between 0.5 and 0.9) were used. Comparison star magnitudes were obtained from the ATLAS catalog \citep{Tonry18}, which is incorporated directly into MPO Canopus. Star subtraction and outright rejection of frames was necessary in cases where the asteroid passed through dense star fields. The photometric aperture ranged from 7 pixels (7.6 arcsec) to 13 pixels (14.2 arcsec).

\subsection{1-m Las Cumbres Observatory Global Telescope Network (LCOGT)}

The Las Cumbres Observatory (LCOGT) is a global network of twenty-five telescopes in three size classes at seven sites around 
the world \citep{Brown2013LCOGT}. For the DART light curve 
observations, the 1.0-m telescope network was used and observations 
were requested using the NEOexchange Target and Observation Manager 
(TOM; \citealt{Lister2021NEOx}) system. Data were obtained from LCOGT 
sites located at:
\begin{itemize}
\item Cerro Tololo Observatory, District IV, Chile (three 1.0-m telescopes; MPC site codes W85, W86, W87)
\item South African Astronomical Observatory, Sutherland, South Africa (three 1.0-m telescopes; MPC site codes K91, K92, K93)
\item McDonald Observatory, Fort Davis, Texas (two 1.0-m telescopes; MPC site codes V37, V39)
\item Teide Observatory, Canary Islands, Spain (two 1.0-m telescopes; MPC site codes Z31, Z24)
\end{itemize}
All the LCOGT 1.0-m images were obtained with the \textit{Sinistro} 
instruments, each containing a 4k$\times$4k Fairchild CCD with 
$15~\mu\textrm{m}$ pixels. The \textit{Sinistro} imagers provide a 
$26.5\mathrm{{}^{\prime}}\times26.5\mathrm{{}^{\prime}}$ field of view 
with an unbinned pixel scale of 
$0.389\mathrm{{}^{\prime\prime}}/\mathrm{pixel}$. All images were 
obtained in $1\times1$ binning mode with a PanSTARRS-w filter 
(equivalent to SDSS $g'+r'+i'$) which provided high throughput between 
400\,nm and 850\,nm. The 
telescopes were tracking at half Didymos's on-sky ephemeris rate 
throughout the observations. Individual exposures times ranged from 
27.5 to 150 seconds.

The reduction of the \textit{Sinistro} images followed a two step 
process. Initial reduction to basic calibrated data products involving 
bias and dark subtraction, flat fielding and astrometric fitting were 
performed automatically within minutes of readout of the frame by the 
LCOGT BANZAI pipeline \citep{McCully2018BANZAI}. The basic calibrated 
data were then automatically retrieved from the LCOGT Science Archive 
 and pipeline processed through the PP 
\citep{Mommert17} and NEOexchange \citep[NEOx]{Lister2021NEOx} 
pipelines. 

Both pipelines use SourceExtractor \citep{Bertin96} to extract 
sources from the image and SCAMP \citep{Bertin06} to perform the 
astrometric registration to the Gaia DR2 catalog \citep{Gaia18} and 
then calibrated against PanSTARRS DR1 catalog \citep{Flewelling20} or 
the Gaia DR2 catalog, depending on the declination of Didymos at the 
time of the observations. This zeropoint calibration within the NEOx 
pipeline was performed using the \texttt{calviacat} 
\citep{Kelley2022-calviacat} package. A preliminary reduction was 
generally done with the PP to perform a curve of growth analysis and an 
optimal aperture radius for the main NEOx reductions and to act as a 
cross-check on the reductions. Due to the low galactic latitude of 
Didymos in the early 2022 October--November data and the variable and 
differential reddening of the field stars, we did not use the features 
of either PP or \texttt{calviacat} to restrict the field stars to 
having solar-like colors. Given the crowded fields, persistence of ejecta, fading of the target, and analysis focused on differential magnitudes, the choice of non-solar type stars for field calibration had no discernible influence on the quality of the photometry calibration.

\subsection{1-m Jacobus Kapteyn Telescope (JKT) \label{subsec:jkt}}

The JKT is equipped with an Andor 2k CCD camera and is situated at the Roque de los Muchachos Observatory on La Palma. The telescope's field of view is 11.6\arcmin~$\times$ 11.6\arcmin, and the image scale is 0.34 arcsec/pixel. The observations were obtained using the Johnson $R$-filter, and we utilised sidereal tracking.  Exposure times of 100s were used.

Data from the JKT were processed using standard reduction procedures and aperture photometry, with the commercial software MPO Canopus following established procedures \citep[e.g.][]{Oszkiewicz20,Oszkiewicz21,Oszkiewicz23}. We selected five comparison stars in each field with significantly higher signal-to-noise ratios than the target, ensuring they had roughly solar colours (approximately $0.5 < B-V < 0.95$ or $0.35 < g-r < 0.85$). An aperture 21 pixels in diameter was employed. For calibration we used the Pan-STARSS DR1 catalog \citep{Flewelling20}. 

\subsection{1-m Swope}

The Swope 1.0-m telescope is located at Las Campanas Observatory, on the Atacama desert in the north of Chile, at an elevation of 2400m. The Swope telescope is equipped with a 4k $\times$ 4k e2v detector with 15 $\mu m$ pixels, covering a 30 $\times$ 30 arcminutes area with 0.435 arcsecond pixels. The Swope dataset encompassed a total of 8733 Sloan-r images, taken across 19 nights. In 5 of 15 nights a single pointing was used to follow Didymos, while for the other nights two pointings were necessary. For each pointing, the brightest 50 stars in the field were selected as standards to achieve photometric calibration of the individual images and to estimate differential photometry of the asteroid.

Swope images are readout by 4 amplifiers, producing 4 quadrant files for each exposure. Each of these quadrants were processed separately with standard techniques, namely bias subtraction, linearity correction, and flat fielding. After normalization by the individual gains, the full image was rebuilt as a single fits file. The astrometric solution was achieved with an iterative process, starting with a preliminary solution created using the WCS routine within the {\it astropy} pacakge, and then improved by matching star positions against their GAIA coordinates. Instrumental aperture photometry was performed using the python package SEP \citep{Barbary18} on every image for the asteroid's and the brightest star's positions, across a set of apertures from 3 to 20 pixels in radius. To estimate the photometric zero points on  individual images, we used several python packages. {\it astroquery} was used to query the VizieR and Horizon databases to identify GAIA sources within 2 arcseconds of our set of bright field stars, and to obtain the coordinates of the asteroid for the given time stamp in each image. The {\it gaiaxpy}\footnote{https://gaia-dpci.github.io/GaiaXPy-website/} python package was used to request and download synthetic photometry of GAIA stars \citep{Gaia21} in Sloan-r band when available. We found more than 30 GAIA stars with available synthetic photometry in most pointings, and in only two cases did we retrieve less than 10 stars. This allowed us to determine robust statistics for the zero points. For each image we estimated a median, rejected outliers, and measured the standard deviation to provide an error on the zero point, which was typically around 0.01-0.02 magnitude per frame. Final photometry of the Didymos-Dimorphos system was estimated adding the zero points to its instrumental magnitude for each image.

\subsection{1-m Zeiss telescope at Tien-Shan Observatory}

The 1 m Zeiss telescope at Tien-Shan Observatory is located at 2800 meters altitude in the Almaty region of Kazakhstan. The observations were carried out with the front illuminated CCD camera PL09000 (made by Finger Lakes Instruments) with a sensor 3056 $\times$ 3056 pixels and a pixel size of 12 $\mu m$. The images covered a 19.1\arcmin~$\times$ 19.1\arcmin~field of view. The asteroid was observed with a Johnson-Cousins $R$ filter. The observations were carried out with the telescope tracking at sidereal rates and with the camera in 2$\times$2 binning mode (producing an image scale = 0.75 \arcsec/pixel). In the end of December, the asteroid was moving across the sky at an angular rate of 1.2 \arcsec/min and thus was trailed by about 2.4 pixels during the 90-second exposures.

Reduction of the images included removal of an average dark frame and normalization with a median dome flat field. Didymos’ brightness was measured with the AstPhot software \citep{Mottola95}. The size of the aperture was chosen to maximize signal to noise based on measurements of several bright stars. An aperture radius of 6 pixels (4.5\arcsec) was determined to be optimal. An elliptical aperture of 6 $\times$ 7 pixels was used for the slightly elongated asteroid. As with the AZT-22 data, these apertures were chosen to roughly approximate isophotes for the stars and asteroid. The $R$ magnitudes of comparison stars were taken from the ATLAS catalog \citep{Tonry18} and were used to calibrate the asteroid using the MPO Canopus software package. 

\subsection{0.9-m Spacewatch}

SPACEWATCH\textsuperscript{\textregistered} operates Steward Observatory’s 0.9-m telescope on Kitt Peak, in Arizona, at an elevation of 2080 m.  Images were obtained with the Spacewatch mosaic camera – a mosaic of four e2v 4k $\times$ 2k CCDs with 13.5 $\mu m$ pixels. It has an effective field of view of 2.9 square degrees at an un-binned pixel scale of 1\arcsec/pixel. The images were obtained un-binned with a broadband Schott OG-515 filter, which has a long-pass transmission profile with a cut-on wavelength at 515 nm. Individual exposures ranged from 16 to 104 seconds across the apparition.
 
The reduction followed standard bias, flat field, and fringe correction techniques. The photometry of the Didymos system was measured and calibrated using the PP \citep{Mommert17} and MPO Canopus following the same procedures as were applied to the LDT and Hall telescopes respectively.

\subsection{0.8-m at the Instituto de Astrofísica de Canarias (IAC80)}

The IAC80 telescope, equipped with the CAMELOT2 instrument, is located at the Observatorio del Teide on Tenerife. CAMELOT2 features a 4k x 4k back-illuminated CCD. The on-sky pixel scale is 0.322 arcsec/pix, providing a theoretical field of view of 22 x 22 arcminutes$^2$. However, due to vignetting caused by the filters, the useful squared field of view is 11.8 x 11.8 arcminutes$^2$. The data were obtained using the Johnson $R$-filter, and sidereal tracking was employed. An exposure time of 135s and aperture diameter of 19 pixels were used.

Data reduction and measurement of photometry for the IAC80 data followed the same procedures as used for data from the JKT (\S\ref{subsec:jkt}).

\subsection{0.7-m AC-32 telescope of the Abastumani Astrophysical Observatory}

The 0.7 m AC-32 telescope is a Maksutov meniscus telescope at the Abastumani Observatory which is located on Konobili Mountain in the Samtskhe-Javakheti region of Georgia at an altitude 1650 m. AC-32 is equipped with a back illuminated 2k $\times$ 2k CCD camera PL4240 (made by Finger Lakes Instruments) with 13.5 $\mu m$ pixels. The camera is installed at the prime focus. AC-32 images a 44.4\arcmin~field of view with an un-binned pixel scale of 1.30\arcsec. We employed a Johnson-Cousins R filter. All AC-32 observations were taken at sidereal rates. The exposures ranged from 120 to 180 seconds, depending on the observing circumstances.

The reduction of AC-32 images was performed using standard dark and flat-field corrections. The average flat-field was calculated as a median of more than seven twilight sky flat images. Aperture photometry was used to measure brightness by means of the AstPhot software package \citep{Mottola95}. An optimal aperture for field stars of 4-5 pixels (5.2 - 6.5\arcsec) was used. An elliptical aperture for the asteroid was extended in the direction of its motion by 1-2 pixels. Again, these aperture dimensions were chosen to roughly approximate isophotes for the stars and asteroid. Relative photometry for the asteroid was performed by subtracting the magnitudes of nearby comparison stars. The comparison stars were chosen with colors close to the Sun. Uncertainties on the instrumental magnitudes of the comparison stars were usually around 0.002-0.005 mag. The comparison stars were calibrated to Johnson-Cousins $R$ band based on reference magnitudes from the ATLAS catalog \citep{Tonry18}.

\subsection{0.65-m Ond\v{r}ejov Telescope}

The 0.65-m telescope is located at Ond\v{r}ejov Observatory, Czech Republic at an elevation of 528~m.
It is operated jointly by the Astronomical Institute of the Academy of Sciences of the Czech Republic 
and the Astronomical Institute of the Charles University Prague, Czech Republic. All images in the DART campaign were obtained with a Moravian Instruments G2-3200 Mk.\,II CCD camera that uses a Kodak KAF-3200ME sensor and a standard Cousins $R$ photometric filter \citep{Bessell90} mounted in the prime focus with a Paracorr coma corrector. The CCD sensor has $2184 \times 1472$ square pixels ($6.8 \mu$m size) with micro-lenses and we used it in $2 \times 2$ binning mode that provided a scale of 1.05~arcsec/pixel and a $19 \times 13$~arcmin$^2$ field of view. For all observations presented here the telescope was tracked at half-apparent rate of the asteroid, providing star and asteroid images of the same profile in one frame.  The asteroid moved slow enough during the observations so that we could use a single set of local reference stars for the observations taken on one night. Individual exposure times ranged from 90 to 180 seconds across the apparition.

The reduction of images followed standard flat field and dark frame correction techniques. The photometry was measured and calibrated in a manner identical to that employed for the 1.54-m Danish telescope. Aperture radii of 4 or 5 pixels (4.2 or 5.25 arcsec) were found optimal for data from the 0.65-m Ond\v{r}ejov telescope.

\subsection{0.64-m at Sugarloaf Mountain Observatory}

Sugarloaf Mountain Observatory is located in South Deerfield, MA, USA, at an elevation of 65 m. The telescope is a 0.64 m reflector. The imager is a SBIG Aluma 3200 CCD using the KAF-3200 chip. This chip contains an array of 2184 $\times$ 1472 pixels unbinned. All images were taken using 2$\times$2 binning. The telescope has a field of view of 23.2 $\times$ 15.6 arc min and a working image scale of 1.27 arcsec/pixel. No filters were used to acquire images and exposure times ranged from 70-100 seconds. Tracking was at the sidereal rate.

All images were processed using dark, bias and flat-field corrections. Image reduction was accomplished using MPO Canopus software. Calculated magnitudes were based on an internal scale using several comparison stars which were selected to be similar to solar color. The magnitudes of the comparisons were those in the $R$ band in the Carlsberg Meridian Catalog  \citep[CMC15, ][]{CMC14}. Measurement apertures were either 11 or 13 pixels in diameter (14.0\arcsec~or 16.5\arcsec) depending on seeing.

\subsection{0.6-m G2 at Star\'a Lesn\'a Observatory}

The 0.60-m f/12.5 Cassegrain telescope is situated near Star\'a Lesn\'a village in Slovakia. It belongs to the Astronomical Institute and is located in the G2 pavilion. For imaging, it uses an FLI CCD camera with 15 $\mu m$ pixels (un-binned). For observing Didymos we used 2 $\times$ 2 binning to produce an effective image resolution of 0.85\arcsec~$\times$ 0.85\arcsec~and a field of view 14.5\arcmin~$\times$ 14.5\arcmin. We used a Johnson-Cousins $R$ filter. The telescope was set to track at sidereal rates so that the asteroid was moving through the field. The exposure time was set to 170 seconds. Light frames were reduced with dark frames and flat fields in a standard manner. For the photometric measurements, we used MaximDL6 with an aperture size of 11 pixels. 

\subsection{0.6-m TRAPPIST North and South}

TRAPPIST-South (TS) is located at the ESO La Silla Observatory in Chile \citep{Jehin11}. TRAPPIST-North (TN) is located at the Ouka\"{i}meden observatory in Morocco. Both TS and TN are robotic 0.6-m Ritchey-Chr\'{e}tien telescopes operating at f/8. TS is equipped with a FLI ProLine 3041-BB CCD camera with a 22\arcmin~field of view and un-binned pixel scale of 0.64\arcsec/pixel. TN is equipped with an Andor iKon-L BEX2 DD camera providing a 20\arcmin~field of view and un-binned pixel scale of 0.60\arcsec/pixel. No binning was used for observations from September to the end of November 2022, after which 2$\times$2 binning mode was used. The Exo filters were used, which are broad blue-blocking filters with a transmission from 0.5 $\mu m$ to the NIR. Exposure times ranged from 45 to 120 seconds.  

The raw images were processed using standard bias, dark and flat fields frames. The photometry was measured using the PP \citep{Mommert17} and calibrated to the $R_c$ Johnson-Cousins band using the PanSTARRS DR1 catalog. Typically, more than 50 field stars with solar-like colors were used in each image. The photometric apertures had fixed radii of 12 pixels for the un-binned observations and 6 pixels for the 2x2 binning mode.

\subsection{0.5-m T72 at Deep Sky Chile Observatory}

The T72 telescope of iTelescope is located in Rio Hurtado Valley, Chile, at an elevation of 1710 m. All T72 images were obtained with the KAF-16200 sensor, a 4500 $\times$ 3600 CCD with 6 $\mu m$ square pixels. KAF-16200 images a 26.93\arcmin~$\times$ 21.53\arcmin~ field of view at an un-binned scale of 0.359\arcsec/pixel. All images were obtained in 2$\times$2 binning mode with a Johnson-Cousins $R$ filter. For all T72 data presented here, the telescope was tracked at sidereal rates, allowing the asteroid to pass through fixed star fields. Individual exposure times were 60 seconds.

The reduction of KAF-16200 images followed standard flat field and bias correction techniques. The images were calibrated by the iTelescope pipeline.  However, the alignment of the images as well as plate solving and photometry, were performed using the software Tycho Tracker. Typically 4 to 7 field stars were used to calibrate each image. Aperture radii ranged from 4 to 8 pixels.

%%%%%%%%%%%%%%%%
% ANALYSIS
%%%%%%%%%%%%%%%%
\section{Lightcurve Analysis} \label{sec:analysis}

The primary objective of the lightcurve campaign was to detect mutual events that could then be used to determine the orbital period of Dimorphos. This required grouping lightcurves into decomposable sets of 1 up to 14 individual lightcurves, where each set was collected over the span of 1 up to a few days. As a rough general rule, coverage across at least two rotations of Didymos ($\sim4.5$ h) outside of mutual events was needed for a successful decomposition. Decomposition sets were defined based on the morphology of the primary lightcurve. Namely, individual lightcurves were added to a decomposition set as long as the morphology of the primary lightcurve remained constant within the signal-to-noise of the data. A new decomposition set was defined when changes to the primary lightcurve were detected. A total of 43 decompositions were performed with 224 individual lightcurves (Table \ref{tab:decomp}). Individual decompositions were given unique IDs indicating the lunation in which the data were obtained followed by an integer indicating the n-th decomposition within the lunation. For example, the third set of decomposed lightcurves in the post-impact L0 lunation was assigned an ID = L0.3. These IDs facilitate mapping of information across Tables \ref{tab:lunations}, \ref{tab:decomp}, and \ref{tab:details}, and the supporting data file that contains all of the lightcurve measurements.

\begin{deluxetable*}{llcccc}[htbp]
\tablecaption{Lightcurve decompositions from the 2022-2023 observations. The columns correspond to the ID for each decomposition, the range of JD dates associated with the data, the number of data points and lightcurves in each decomposition, the order of Fourier series used to fit the primary lightcurve, and the rms associated with those fits. \label{tab:decomp}}
\tabletypesize{\footnotesize}
\tablehead{\colhead{ID} & \colhead{JD Range} & \colhead{Points} & \colhead{Lightcurves} & \colhead{Fit Order} & \colhead{rms}}
\startdata
pre-L1  &   2459762.6669 - 2459767.9728   & 706 & 5   & 9    & 0.0067 \\
pre-L2  &   2459791.6504 - 2459791.9057   & 125 & 1   & 10    & 0.0080 \\
pre-L3.1  &   2459809.5313 - 2459810.7136   & 229 & 2   & 12  & 0.0132 \\
pre-L3.2  &   2459813.5676 - 2459817.5414   & 2135 & 7  & 9    & 0.0097 \\
pre-L3.3  &   2459821.5357 - 2459824.9126   & 2123 & 4  & 9    & 0.0084 \\
pre-L3.4  &   2459826.5995 - 2459828.6771   & 286 & 3   & 9   & 0.0066 \\
pre-L4.1  &   2459834.5636 - 2459838.9102   & 1031 & 6  & 9    & 0.0091 \\
pre-L4.2  &   2459840.8258 - 2459844.9064   & 1248 & 7  & 11   & 0.0066 \\
pre-L4.3  &   2459845.7830 - 2459848.8959   & 2610 & 9  & 9    & 0.0070 \\
L0.1      &   2459850.6063 - 2459850.8867   & 577 & 2   & 7    & 0.0066 \\
L0.2      &   2459851.6116 - 2459851.8663   & 1284 & 3  & 10    & 0.0053 \\
L0.3      &   2459852.5897 - 2459852.8888   & 1675 & 6  & 11   & 0.0058 \\
L0.4      &   2459853.4068 - 2459853.8951   & 1362 & 5  & 12   & 0.0044 \\
L0.5      &   2459854.4089 - 2459854.9549   & 1595 & 5  & 11   & 0.0052 \\
L0.6      &   2459855.4172 - 2459855.7280   & 1238 & 6  & 11   & 0.0046 \\
L0.7      &   2459856.5925 - 2459856.8663   & 657 & 4   & 11  & 0.0053 \\
L0.8      &   2459857.4693 - 2459857.7773   & 774 & 5   & 10   & 0.0066 \\
L0.9      &   2459858.4297 - 2459858.8509   & 1214 & 5  & 9    & 0.0057 \\
L0.10      &   2459859.6737 - 2459859.8807   & 712 & 3   & 9   & 0.0062 \\
L0.11     &   2459860.4382 - 2459860.8954   & 805 & 3   & 10  & 0.0070 \\
L0.12      &   2459861.4424 - 2459861.8936   & 837 & 4   & 9   & 0.0069 \\
L0.13      &   2459862.6761 - 2459862.7933   & 767 & 3   & 12  & 0.0083 \\
L1.1      &   2459869.6181 - 2459870.0207   & 165 & 2   & 9   & 0.0100 \\
L1.2      &   2459872.8759 - 2459874.0173   & 374 & 3   & 11  & 0.0085 \\
L1.3      &   2459876.8377 - 2459880.0287   & 666 & 8   & 9   & 0.0094 \\
L1.4      &   2459880.5413 - 2459886.0129   & 1455 & 14  & 14   & 0.0094 \\
L2.1      &   2459900.7059 - 2459901.7576   & 715 & 4  & 9   & 0.0065 \\
L2.2      &   2459901.8103 - 2459904.0346   & 986 & 6  & 10  & 0.0074 \\
L2.3     &   2459904.5753 - 2459907.0441   & 1019 & 9  &12    & 0.0067 \\
L2.4      &   2459907.7170 - 2459909.0380   & 817 & 8   & 13  & 0.0077 \\
L2.5      &   2459909.7477 - 2459911.5513   & 492 & 5   & 13  & 0.0072 \\
L2.6      &   2459911.4180 - 2459913.9069   & 1176 & 10  & 12   & 0.0084 \\
L2.7      &   2459914.6155 - 2459915.9442   & 647 & 6  &12   & 0.0065 \\
L3.1      &   2459927.9180 - 2459929.0688   & 713 & 2  & 14  & 0.0050 \\
L3.2      &   2459930.4872 - 2459931.0263   & 300 & 3  & 14  & 0.0064 \\
L3.3      &   2459932.4297 - 2459934.9537   & 1035 & 10  & 12   & 0.0091 \\
L3.4      &   2459935.2858 - 2459937.5623   & 448 & 4  & 13  & 0.0093 \\
L3.5      &   2459938.0873 - 2459940.3691   & 1036 & 7  & 13   & 0.0101 \\
L3.6      &   2459941.4396 - 2459944.8952   & 415 & 4  & 11  & 0.0060 \\
L4.1      &   2459955.5764 - 2459962.7933   & 501 & 6  & 11  & 0.0082 \\
L4.2      &   2459968.5436 - 2459974.7416   & 1058 & 10 & 12   & 0.0098 \\
L5.1      &   2459986.5711 - 2459986.9064   & 255 & 1  & 11  & 0.0079 \\
L5.2      &   2459992.6056 - 2460000.8285   & 269 & 4  & 11  & 0.0059 \\
\hline
Totals:      &      &   38532 & 224  &  \\
\enddata
\end{deluxetable*}

Observers on the investigation team submitted lightcurves for decomposition as simple ascii files, typically containing JD, magnitude, and magnitude error. Our methodology for decomposing these lightcurves into their constituent parts can be summarized with the following steps, which are then described in more detail in the remainder of this section: 
\begin{itemize}

    \item The observed JD values were light time corrected to be in the reference frame of the asteroid.

    \item For each individual lightcurve, magnitudes were differentially corrected for small changes in geometry (phase angle, geocentric range, heliocentric range) within the night.
    
    \item Based on the latest orbit solution of Dimorphos, data taken within mutual events were masked out.
        
    \item Lightcurves were converted to differential magnitudes by subtracting off the mean magnitude outside of the masked mutual events. This zero point offset was included as a fit parameter for each individual lightcurve.
    
    \item A linear trend was fit to each lightcurve to correct for time variable brightness of  ejecta within the photometric apertures.

    \item The primary rotational signatures in the differential, corrected lightcurves were fit with Fourier series based on data points outside of mutual events.

    \item The Fourier fits were subtracted off of the differential lightcurves to generate residuals that isolated the mutual events. 

    \item The root-mean-square (rms) of the residuals outside of mutual events was calculated to assess the quality of each decomposition.
    
\end{itemize}

\begin{figure}[h!]
    \centering
    \includegraphics[width=\textwidth]{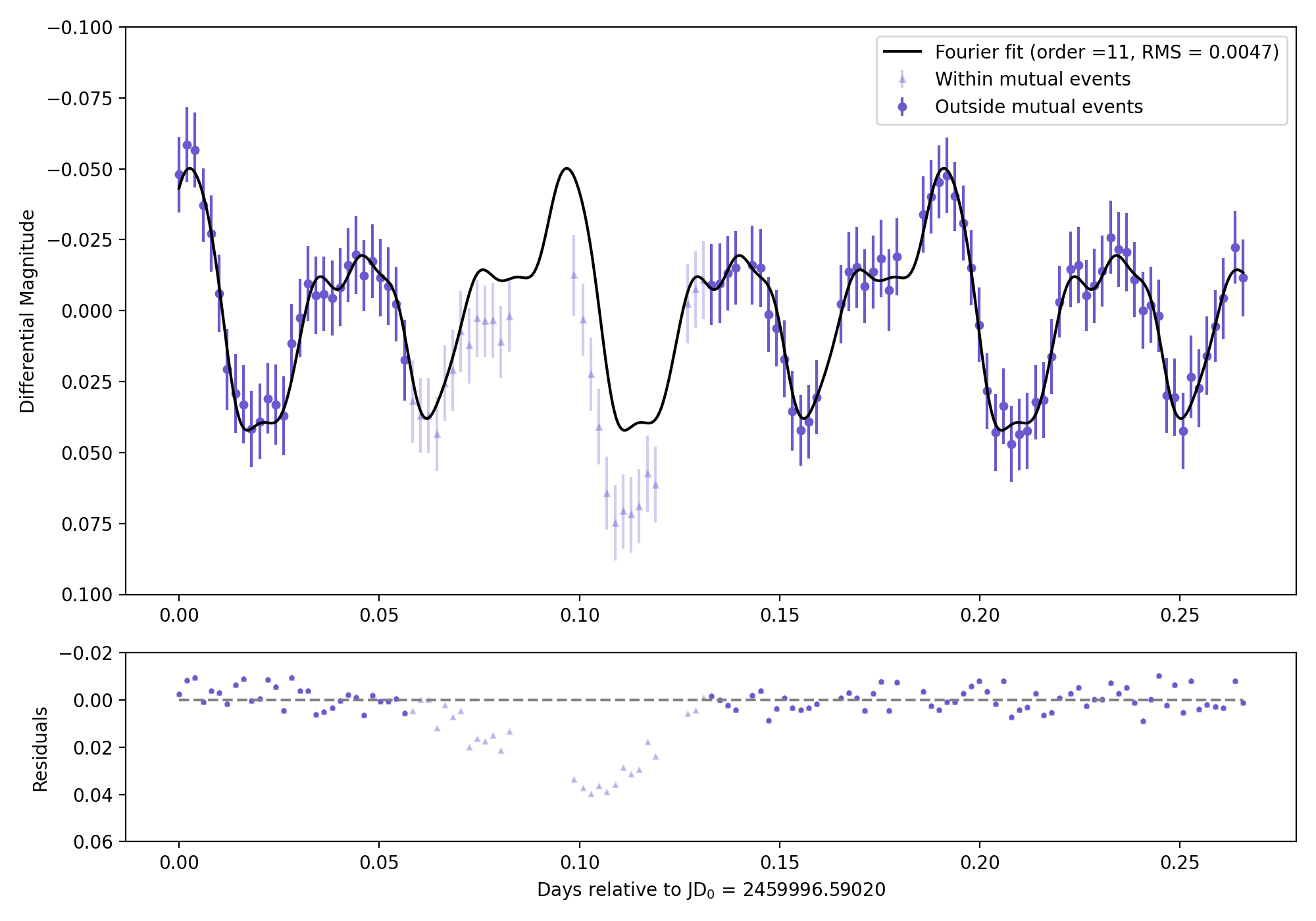}
    \caption{Light time and geometry corrected differential lightcurve measured on 21 February 2023 from the LDT (top). This lightcurve was part of the L5.2 decomposition. The Fourier fit to the differential lightcurve was subtracted off the measured signal to produce residuals that isolated a primary eclipse and occultation centered around t = 0.1 days (bottom). The low rms of the fit (=0.0047) and relatively flat residuals outside of the mutual event are indicators of a clean decomposition. The reported photometric error bars in the top panel were ignored when fitting the Fourier series.}
    \label{fig:ldt_lightcurve}
\end{figure}

An example of a single decomposed lightcurve from the LDT on 21 February 2023 is shown in Figure \ref{fig:ldt_lightcurve}.

The decomposition process began with light time correction so that all measurements were tied to the reference frame of the asteroid. We adopted a simple first order (non-iterative) light time correction by subtracting Didymos' topocentric distance divided by the speed of light from the JD times of observation. Iterative corrections accounting for the motion of Didymos during this light travel time interval were not applied. These first order corrections ranged from about 4 minutes down to 30 seconds at closest Earth approach in early October. Based on the motion of Didymos, subsequent iterative corrections would have been at least three orders of magnitude smaller and thus insignificant relative to individual exposure times and features in the lightcurve.

Within each observing session, the viewing geometry -- characterized by solar phase angle, geocentric range, and heliocentric distance -- changed slightly, which caused small monotonic changes in brightness. We corrected for this by retrieving with the Python \emph{astroquery} package an ephemeris for Didymos from JPL Horizons. Horizons calculates ephemerides using the HG magnitude system \citep{Bowell89} and used values of $G=0.15$ and $H=18.12$ for Didymos. We used this ephemeris to compute differential changes in magnitude relative to the mean. Those differential changes were applied to each measured lightcurve independently. Though we show in Section \ref{sec:mags} that an HG model does not provide the best fit to the entirety of the post-impact photometry, these geometry-dependent magnitude corrections were so small ($<0.1$ mag for each lightcurve) that differences in photometric models or specific HG parameters were insignificant to the decomposition process.

To properly fit the rotational signature of Didymos, we had to identify and mask out portions of each lightcurve that were taken during mutual events (Figure \ref{fig:ldt_lightcurve}). This masking was critical to ensure that the Fourier fit to the rotational lightcurve of Didymos was not affected by mutual events. Defining masks was straightforward prior to impact, because the orbit of Dimorphos was well constrained \citep{Naidu22,Scheirich22}. After DART impact a new orbit solution was found within the first 2 days \citep{Thomas23}, again allowing for identification of mutual events. However, the post-impact orbit solution continued to get refined throughout the apparition. Our mutual event masks thus reflected the most current orbit solution with an additional buffer of up to 0.025 day (36 min) at the beginning and end of each event to account for uncertainties in event predictions. With these masks defined we converted our lightcurves to differential magnitudes by subtracting off mean magnitudes outside of mutual events.

A final correction was applied to each lightcurve to account for the changing brightness of impact ejecta within photometric apertures. Originally this was conceived as a way to compensate for fading of ejecta as it escaped the system. These were small corrections, applied to account for measured ejecta fading rates of $\sim0.1$ mag/d \citep[e.g.][]{Graykowski23}. So for a typical lightcurve, this involved a linear correction of no more than a few hundredths of a magnitude across the observing session. However, given the heterogeneity of the overall data set, it was not possible to disentangle this slope correction from other observational issues such as variable extinction or seeing effects. Furthermore, it was found that the ejecta were not simply fading in a monotonic fashion. Instead, there is clear evidence of complex variability in ejecta brightness. For example secondary collisions may have produced an increase in ejecta about 8 days after impact \citep[see \S\ref{sec:mags} and ][]{Kareta23}. As such we allowed the slope correction to vary from -0.25 to +0.25 mag/d for all data sets. This decision was supported by lower rms values in the final decompositions when these slope corrections were applied. 

With the differential, fully corrected lightcurves, we fit the data for each decomposition with a Fourier series of the form:
\begin{equation}
    m(t) = \sum_{n=1}^{k}\left[A_n cos\frac{2\pi n}{P}(t-t_0) + B_n sin\frac{2\pi n}{P}(t-t_0)\right]-\beta (t-t_0) + \delta m
\label{eqn:lc}    
\end{equation}
where $m(t)$ is the differential magnitude at time $t$, the sum over $n$ defines the order $k$ of the Fourier series, $A_n$ and $B_n$ are the Fourier coefficients, $P$ is the rotation period of Didymos \citep[=2.2600 hr, ][]{Pravec06}, $t_0$ is the start time of the lightcurves included in each decomposition set, $\beta$ is the slope parameter in the range -0.25 to 0.25 mag/d correcting for monotonic variability of the ejecta, and $\delta m$ is a small differential offset applied to each lightcurve individually to minimize the rms in the final decomposition. This formalism is similar to previous works \citep[e.g.][]{Pravec00,Pravec06,Pravec22}, but does differ slightly. For simplicity we fit the measured magnitudes as opposed to converting and fitting units of flux. We found that the quality (rms) of Fourier fit was identical when fitting fluxes or magnitudes. Fitting in logarithmic (magnitude) space was likely adequate for these data because of the relatively low amplitude of the lightcurves.  We also did not explicitly fit a mean magnitude term in Equation \ref{eqn:lc}. This is because this equation was applied to differential lightcurves normalized by their individual means and offset with the small $\delta m$ corrections, typically $<0.01$ mag.  Given the heterogeneity of our data set in terms of filters and calibration techniques, fitting to a mean magnitude would not have been beneficial. We also did not include a term for the rotational signature from Dimorphos. That analysis is saved for future work.

A ``curve of growth" approach was employed to determine the optimal Fourier order (Table \ref{tab:decomp}). A range of orders $k$ from 5 to 15 were scanned, with the optimal fit corresponding to the $k$ value for which the rms on individual lightcurves changed by $<0.001$ mag. The period of the primary $P$ was held constant throughout the apparition as there was no clear evidence that the rotation period of Didymos was changed significantly by the DART impact.

The final residuals and rms values were determined by subtracting the Fourier fit from all lightcurves in a decomposition set. The rms was computed based on residuals outside of mutual events (Figure \ref{fig:ldt_lightcurve}). Attempts to compute formal chi-squared statistics were less useful than rms because of the inconsistency in which error bars were reported on the measured photometry. Some data sets had significantly over-estimated errors, while others were significantly under-estimated. Lacking a way to homogenize error bars, we simply ignored them in the fitting process and computed the RMS residuals to assess the data quality of individual lightcurves.

The process to generate sets of decomposed lightcurves was highly iterative, involving regular adjustments when new data were added or reductions were updated. The overall rms residuals as well as the rms values associated with individual lightcurves were used to reject or accept data. On average, rms residuals $<0.015$ mag were required for acceptance. Exceptions were made, for example, if a noisier lightcurve covered key rotational phases not represented by other lightcurves. Generally, the range of dates for each decomposition set were defined when changes in the morphology of the primary lightcurve exceeded $\sim0.015$ mag. This led to many individual decompositions around the time of closest Earth approach (lunation L0) when the viewing geometry to the system changed fastest, and longer decomposition intervals at the beginning and end of the apparition (Table \ref{tab:decomp}). Other factors that were included in this iterative process included the order of the Fourier series, the start and end times of mutual events (based on ephemeris updates), the zero point offsets for each individual lightcurve, and the slope parameter used to account for time-variable ejecta.

%%%%%%%%%%%%%%%%
% DECOMPOSITION
%%%%%%%%%%%%%%%%
\section{Decomposition Results} \label{sec:results}

The full suite of lightcurve data and decompositions are included with this manuscript as a supplementary data file. A representative subset of these data are shown in Table \ref{tab:data}. The columns in this file are: light time corrected Julian Date, measured apparent magnitude, differential magnitudes which have been geometry corrected, slope corrected, mean subtracted, and zero point offset (\S\ref{sec:analysis}), decomposed residuals which are the differential magnitudes with the Fourier fits subtracted off, and IDs for the observing runs and decompositions. The combination of run IDs and decomposition IDs (Table \ref{tab:decomp}) provide a unique mapping to each individual lightcurve. The run ID indicates the observatory and UT date associated with the start time of each lightcurve, and map to the observational details in Table \ref{tab:details}. In some cases an observatory may have contributed two lightcurves from the same UT date (e.g. LCOGT data from 2022-09-12), in which case the letters `a' and `b' were appended to the Run IDs to distinguish those as distinct. The full suite of data in the supplementary file are organized in chronological order with each lightcurve presented as an uninterrupted block. This data file does not include an indication of which data points were measured within mutual events. Computed beginning and end times for mutual events based on the two independent orbital models \citep{Naidu23,Scheirich23} are included with those publications. Those event time predictions can be combined with the data provided here for future analyses.

\begin{deluxetable*}{lcccrr}[!h]
\tablecaption{Data from the DART lightcurve campaign \label{tab:data}}
\tablehead{\colhead{JD} & \colhead{Mag.} & \colhead{Differential mag.} & \colhead{Residual mag.} & \colhead{Run ID} & \colhead{Decomposition ID}}
\startdata
2459762.66372 & 18.752 & -0.010 & 0.0051 & Magellan\_2022-07-02 & pre-L1 \\
2459762.66520 & 18.764 & 0.002 & 0.0104 & Magellan\_2022-07-02 & pre-L1 \\
2459762.66661 & 18.754 & -0.008 & -0.0017 & Magellan\_2022-07-02 & pre-L1 \\
2459762.66955 & 18.742 & -0.019 & -0.0018 & Magellan\_2022-07-02 & pre-L1 \\
2459762.67097 & 18.763 & 0.002 & 0.0290 & Magellan\_2022-07-02 & pre-L1 \\
...
\enddata
\tablecomments{This table is included in its entirety as a supplementary data file in machine-readable format. A portion is shown here for guidance regarding form and content.}
\end{deluxetable*}

\begin{figure}[h!]
    \centering
    \includegraphics[width=\textwidth]{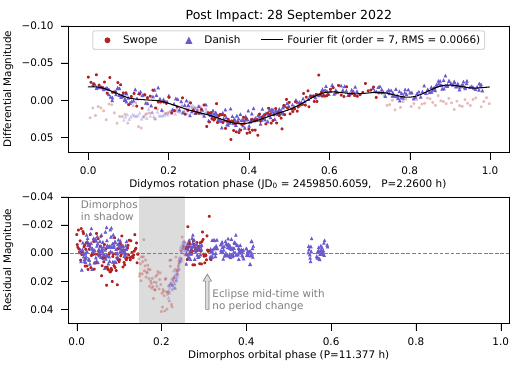}
    \caption{The first mutual event after DART impact was detected in lightcurves from 28 September 2022, only $\sim29$ hours after impact. This event was observed by the Swope (red) and Danish (blue) telescopes. Faded points correspond to data taken within the mutual event. Decomposition of these lightcurves (bottom) revealed a secondary eclipse clearly offset in time relative to an unperturbed orbit solution (gray arrow). This offset of about 1 hour, measured after Dimorphos completed two post-impact orbits, provided the first indication in lightcurves for the 33 minute orbit period change \citep{Thomas23}.}
    \label{fig:sep28}
\end{figure}

The decomposed residuals from the 2022-2023 campaign provided a foundation for determining the pre- and post-impact orbit of Dimorphos \citep{Thomas23}. Immediately following DART impact, the expectation was that mutual events might not be detectable with lightcurves for days or even weeks due to obscuration by ejecta \citep{Fahnestock22}. Fortunately, continued observations during this time revealed the first post-impact mutual event just $\sim29$ hours after impact (Figure \ref{fig:sep28}). A few assumptions facilitated the identification of this event. First, the head-on geometry of the spacecraft impact suggested that the orbit period of Dimorphos would decrease from its pre-impact value of 11.92 h. Second, models predicted likely values for $\beta$ of 1-5 \citep{Stickle22}, which translated to a period change of about 10 min up to 1 h \citep{Meyer21}. Third, the known geometry of the system \citep{Naidu22,Scheirich22} suggested that secondary eclipses in late September would last for about an hour. Based on these expectations, we attempted decomposing the immediate post-impact lightcurves with mutual event masks that spanned a plausible range of new orbital periods. In particular we applied a 1 hour mask across a range of orbit periods from 11 to 12 hours in steps of 0.1 h. This period scan revealed a secondary eclipse in data from UT 28 September that corresponded to a new orbit period of around 11.4 h (Figure \ref{fig:sep28}). The depth of this mutual event was about 0.03 mag, or 60\% of that predicted by the models of \citet{Scheirich22}. This suggested, unsurprisingly, that the increased flux from residual ejecta muted the depth of this mutual event. In Section \ref{sec:depth} we explore how the depth of mutual events evolved as ejecta cleared out of the system.

\begin{figure}[h!]
    \centering
    \includegraphics[width=\textwidth]{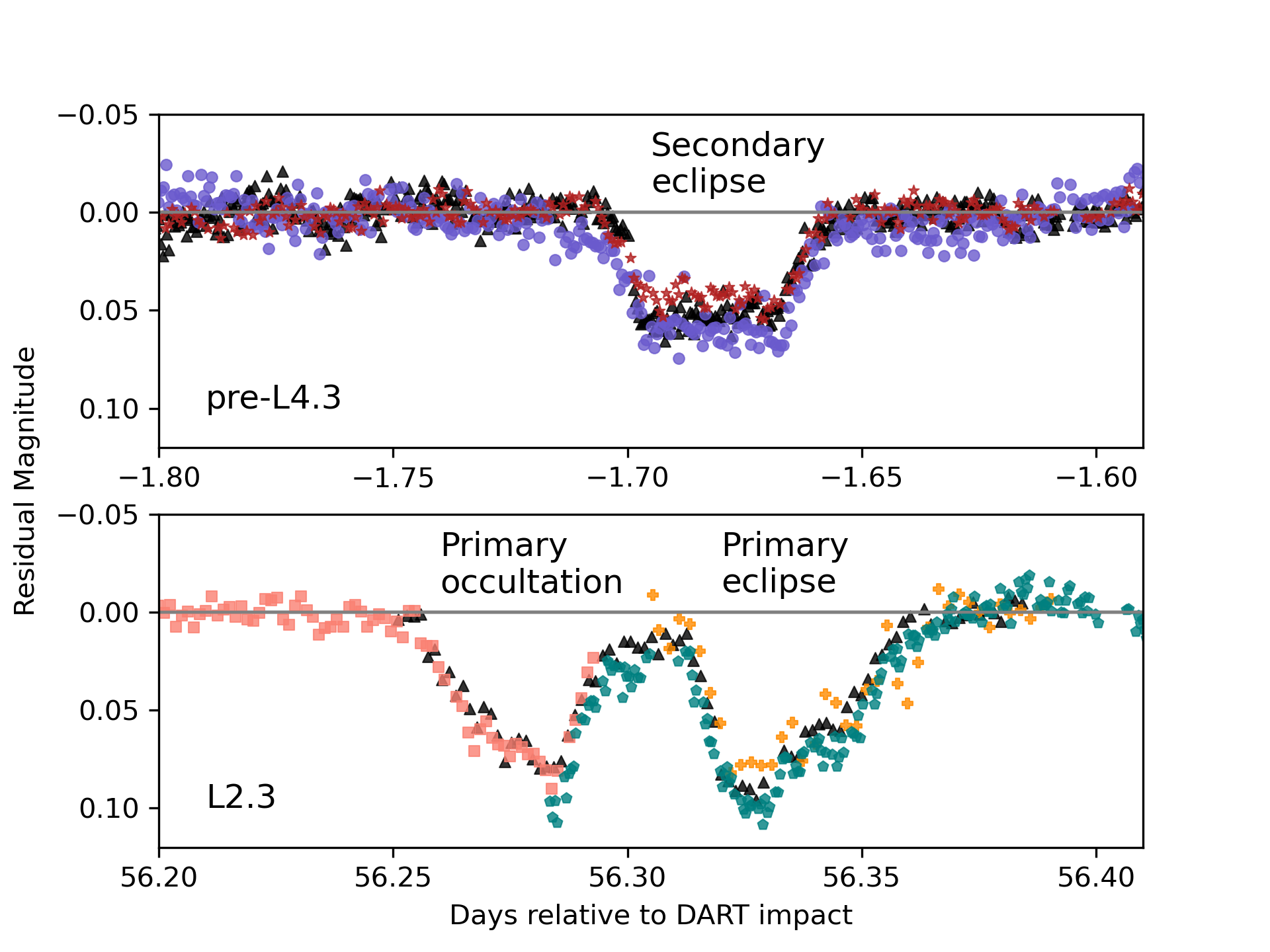}
    \caption{Examples of secondary (top) and primary (bottom) mutual events that were simultaneously sampled by more than one observatory. The top and bottom panels correspond to the pre-impact L4.3 and post-impact L2.3 lunations respectively. These data demonstrate the high precision achieved with these observations, generally 0.01 magnitude or better, and the good agreement across facilities and reduction methods. The key for these points is: 1-m Swope = blue circles, 0.6-m TRAPPIST = red stars, 1.5-m Danish = black triangles, 1-m LCOGT = salmon squares, 1.1-m Hall = orange pluses, and 2.4-m MRO = teal pentagons. Visualization of the system at specific mutual event geometries (e.g. onset of primary occultation) are shown in \citet{Naidu23}.}
    \label{fig:resid}
\end{figure}

A wide variety of mutual events were seen as the viewing geometry of the Didymos system changed throughout the 2022-2023 apparition. This variety included discrete (separated in time) primary and secondary eclipses and occultations, as well as events that overlapped in time. In some cases events were covered simultaneously by multiple observatories (Figure \ref{fig:resid}). These simultaneous observations provided important consistency checks across facilities and data reduction methods. Typically, the data were consistent at the level of $\sim0.01$ magnitude. Comparing simultaneous observations also helped to highlight sections within individual data sets that may have had reduction problems and thus could not be accepted as part of the final data set.

In general, these decomposed residuals provided the best means for validating and accepting individual lightcurves. The full lightcurve data set demonstrated high data quality (e.g. accuracy and precision $\sim0.01$ magnitudes) sustained over an 8 month observing window, with some individual lightcurves meeting these standards for more than 8 hours in a single night (Table \ref{tab:details}).

%%%%%%%%%%%%%%%%
% EVENT DEPTHS
%%%%%%%%%%%%%%%%
\section{Mutual event depths} \label{sec:depth}

With nearly 8 months of data spanning the impact apparition, we assess the evolution of mutual events relative to models. Such models, e.g., \citet{Naidu22,Scheirich22}, have been well demonstrated to reproduce the timing of mutual events. Here we focus on the predicted depths of mutual events relative to the data. We approach this comparison cautiously because these models were not explicitly developed to match the detailed shapes of mutual events. Factors such as topography, non-uniform albedo, or photometric scattering properties could all have contributed to discrepancies between the data and model. We discuss these complications further in Section \ref{sec:disc}.

\begin{figure}
    \includegraphics[width=0.5\textwidth]{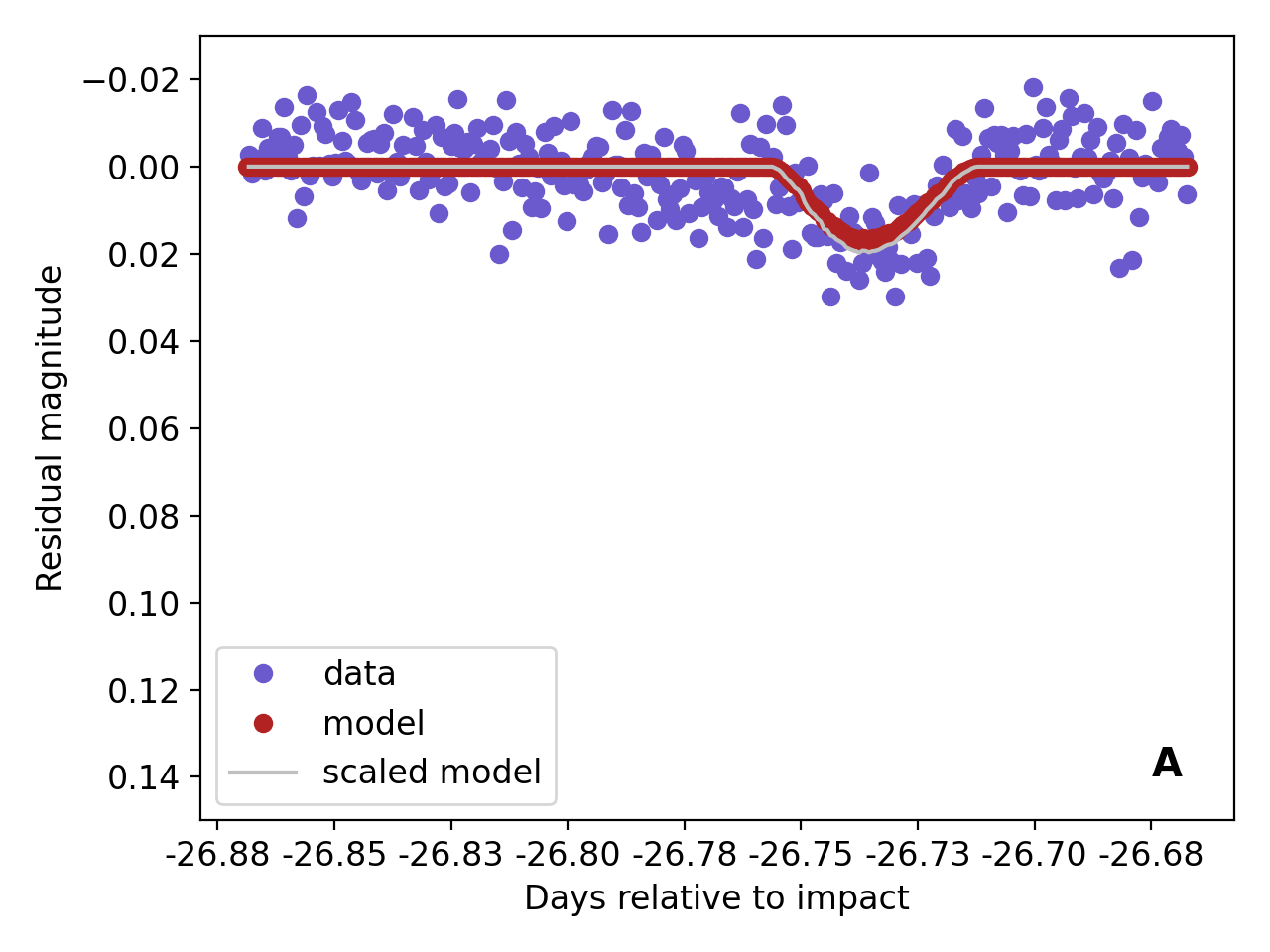}
    \includegraphics[width=0.5\textwidth]{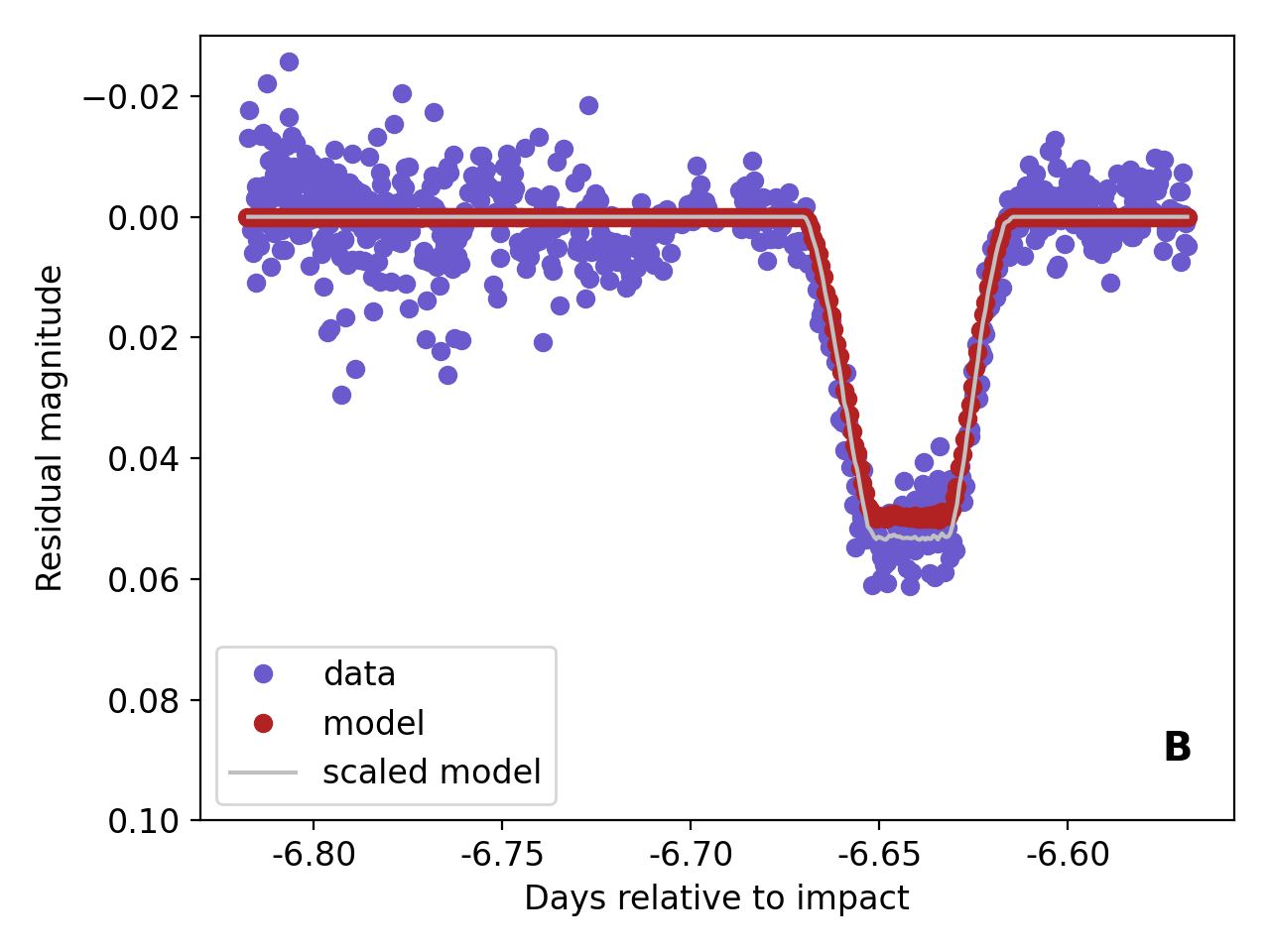}
    \includegraphics[width=0.5\textwidth]{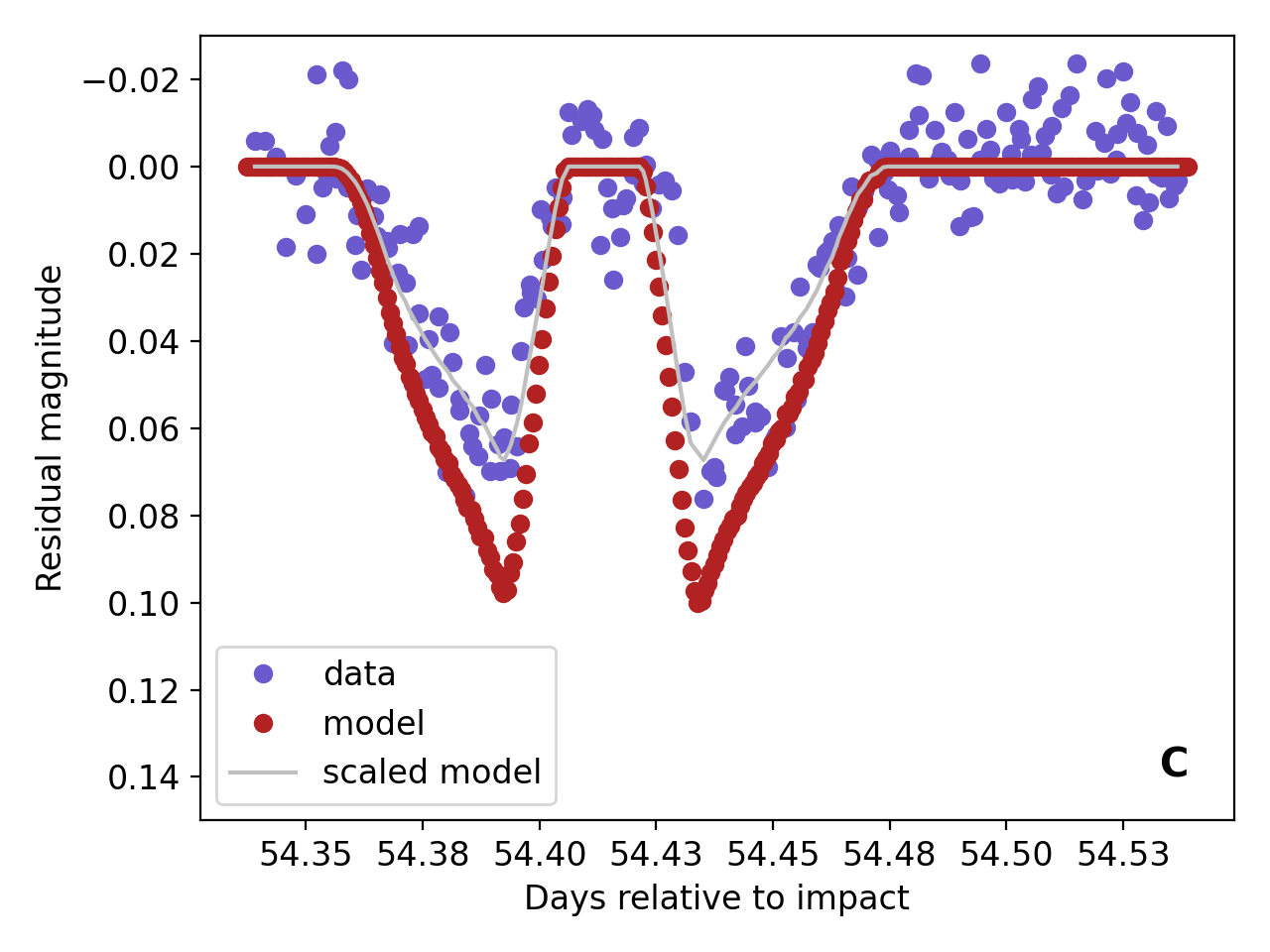}
    \includegraphics[width=0.5\textwidth]{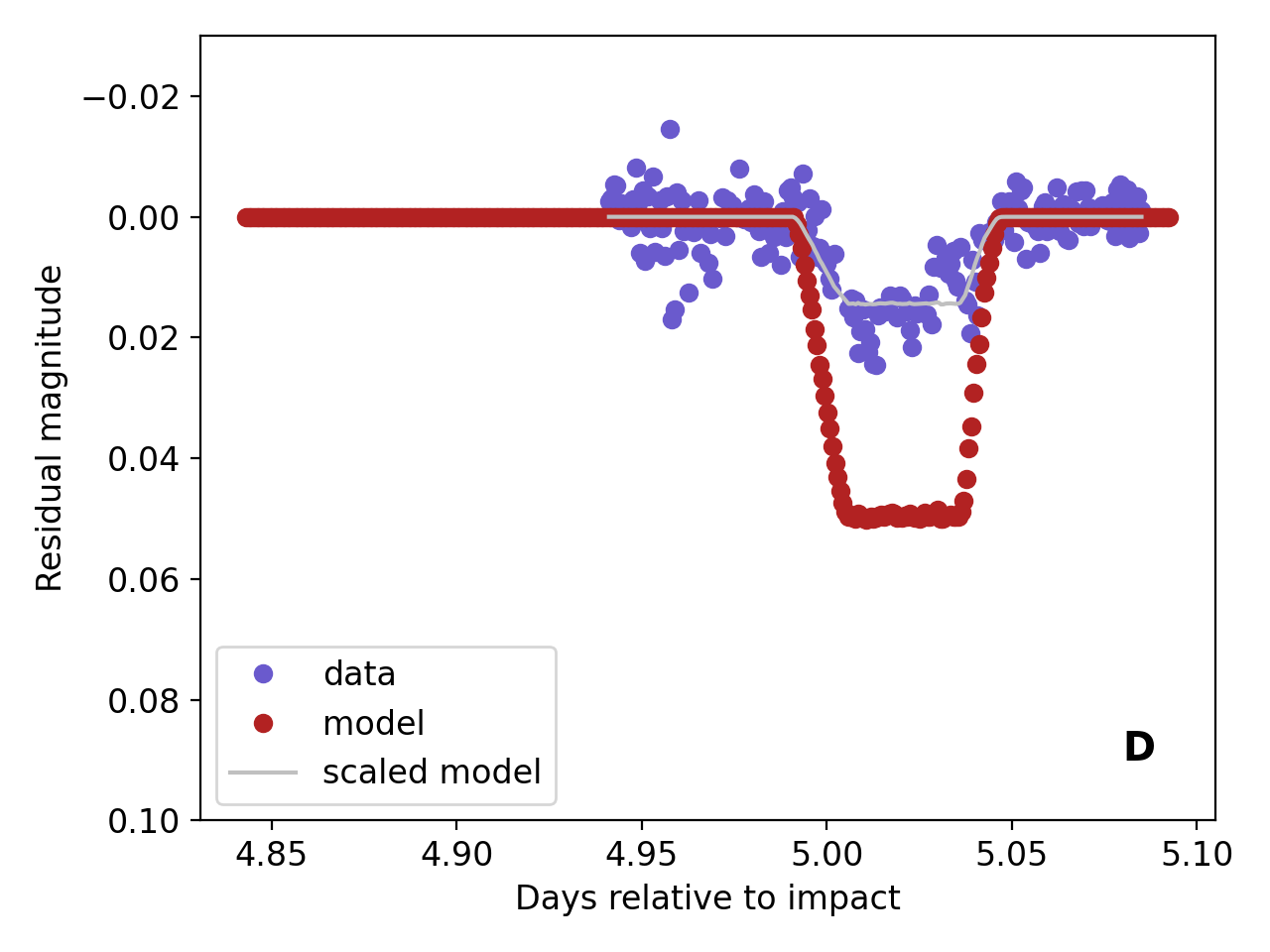}
    \caption{Select mutual events from the pre (panels A and B) and post (panels C and D) impact data. Panel A is a primary eclipse, panel B a secondary eclipse, panel C shows both a primary eclipse and primary occultation, and panel D is a secondary eclipse. The mutual event models of \citet{Naidu23} are over-plotted. In all cases the models are multiplicatively scaled to determine a best-fit to the data. In the post-impact data (C, D) the models over-predict the depth of mutual events, a difference that can be attributed to the affects of residual ejecta in the system. Visualization of the bodies in specific mutual event configurations are presented in \citet{Naidu23}.}
    \label{fig:models}
\end{figure}

The details of the photometric model used for this analysis are presented in \citet{Naidu23}. This model was primarily developed to facilitate improved measurement of mutual event times from decomposed lightcurves and thus served the mission's Level 1 requirement of measuring the period change. This model used rotationally symmetric ellipsoid shapes for Dimorphos and Didymos based on the extents reported by \citet{Daly23}, though the dimensions of Dimorphos were scaled up by 10\% to calibrate against pre-impact mutual event data. The photometric model used the latest orbit solution for Dimorphos \citep{Naidu23}, and for simplicity treated the net photometric signature of the system assuming the Lommel-Seeliger (LS) law for diffuse scattering \citep[e.g.][]{Kaasalainen01}. LS parameters were adopted to be representative of S-type asteroids \citep{Huang17}. This model produced realistic predictions for both the timing and morphology (shape, depth) of mutual events. Figure \ref{fig:models} shows this model as compared to select mutual events in pre- and post-impact lightcurves. In these cases it is clear that the timing and qualitative shape of the events are well represented by the models. However, the depths of events in the post-impact data are sometimes shallower than model predictions. This can be attributed to the influence of residual ejecta in the system. 

With ejecta present, the observed depth of mutual events in magnitudes can be expressed as:
\begin{equation}
   m_{d,obs} = -2.5~log\left(\frac{(f_{P}+f_{s}-\Delta f)~e^{-\tau}+f_{e}}{(f_{P}+f_{s})~e^{-\tau}+f_{e}}\right) 
\label{eqn:event_depth}    
\end{equation}
where $f$ is the flux from the primary Didymos ($P$), secondary Dimorphos ($s$), or ejecta ($e$), the change in brightness due to a mutual event is $\Delta f$, and extinction from the surrounding ejecta is characterized by attenuation of flux for optical depth $\tau$. The modeled mutual event depths ($m_{d,model}$) did not include the extinction ($e^{-\tau}$) or ejecta ($f_e$) terms in Equation \ref{eqn:event_depth}. Thus when the ejecta is optically thick (i.e. $\tau\sim1$) or is contributing significant flux, we expect the model and data to show significant differences.

We estimate the difference in magnitude between the model and observations, $\Delta M = m_{d,model} - m_{d,obs}$, by minimizing the rms between the data and scaled versions of the model (e.g. Figure \ref{fig:models}). This was done for all mutual events in the data set. For each event, we scanned a range of multiplicative scaling factors from 0.01 to 2.0 in steps of 0.01 to find the minimum rms. For each mutual event, the value for $\Delta M$ was then the difference in minimum brightness (maximum magnitude) of the nominal model relative to the scaled model. With this approach we determined that the first mutual events after impact were $\sim0.04$ magnitudes shallower than predicted by the models. If we assume (incorrectly) that this difference is exclusively due to extinction by the surrounding ejecta, i.e.  flux from the ejecta $f_e$ can be ignored and that the observed flux is simply $f_{model}\times e^{-\tau}$, then standard relationships between optical depth and extinction suggest $\tau = \frac{\Delta M}{1.086} \sim 0.04$. It is thus clear that the ejecta became optically thin within the first day or so after impact.

For a period of about 6 weeks in late 2022 (lunations L1 and L2), when Didymos was at relatively high solar phase angles $\alpha$, eclipses and occultations were separated in time (Figure \ref{fig:models}, panel C). In our data this separation was apparent from October 24 ($\alpha=76^\circ$) to December 2 ($\alpha=46^\circ$). In this window, we were able to independently assess the model fit to occultations and eclipses.  Outside of this window, occultations and eclipses overlapped in time, and we thus computed a single best-fit scaling factor that did not consider the events separately.

\begin{figure}[h!]
    \centering
    \includegraphics[width=0.49\textwidth]{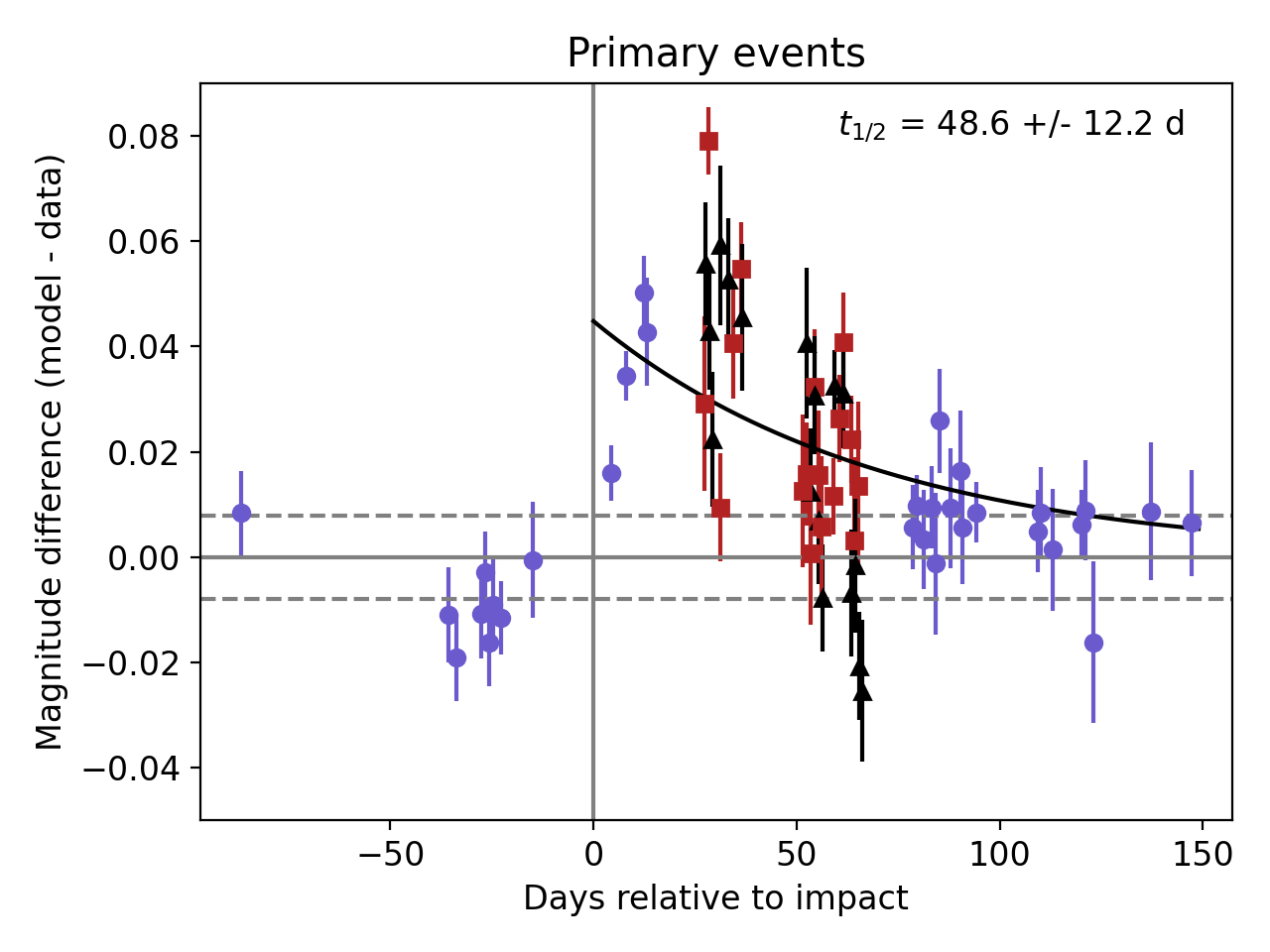}
    \includegraphics[width=0.49\textwidth]{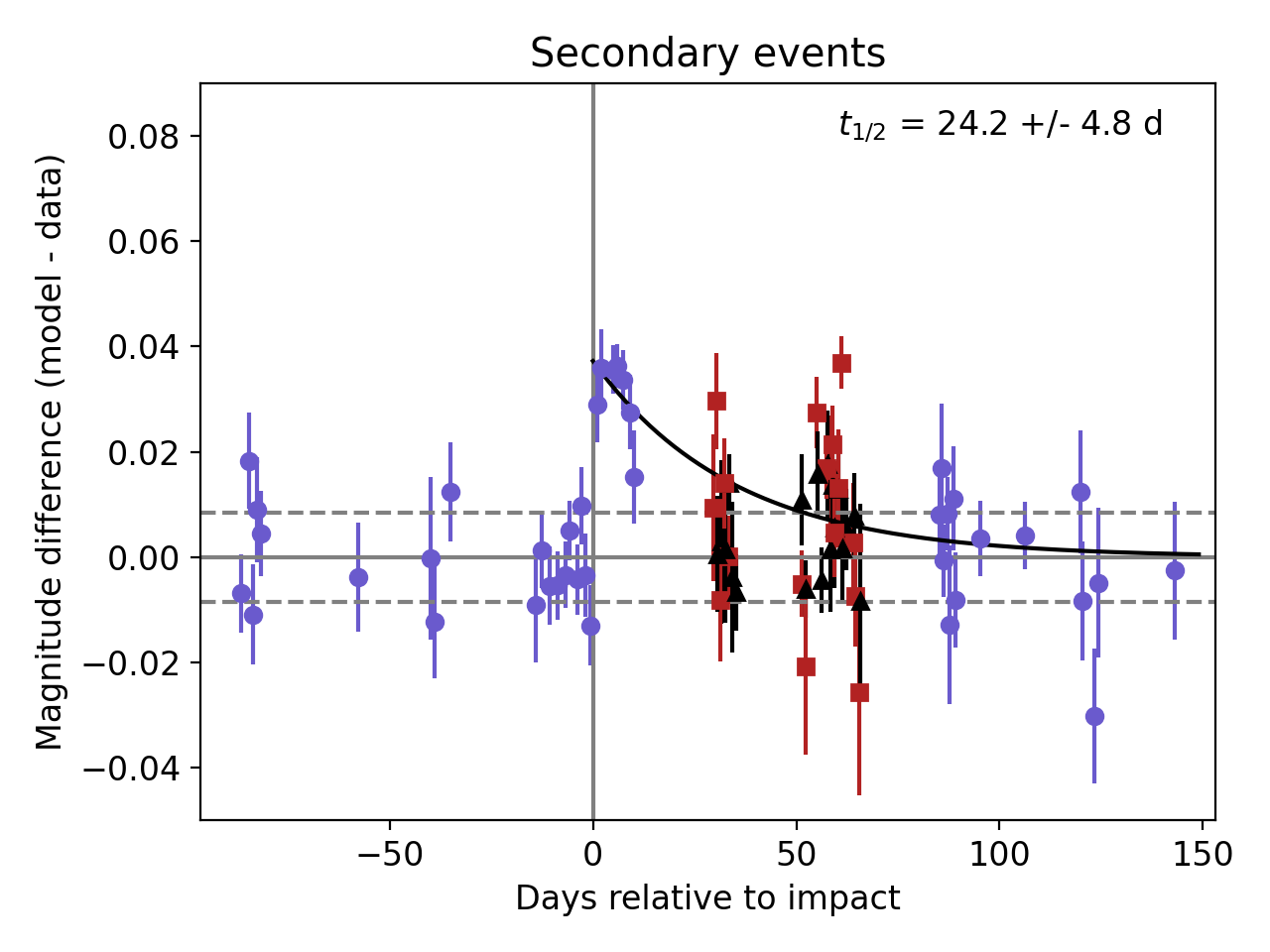}
    \caption{Difference in magnitude $\Delta M$ between modeled and observed mutual event depths. The events prior to impact provided a baseline for the expected variance, i.e. how well the model represented the data. Increased positive differences after impact (vertical line) are due to the influence of ejecta on the observations. The horizontal dashed lines are $\pm1\sigma$ of the pre-impact values. The error bars on the points are the rms associated with the scaled models that best fit the data. Exponential decay curves (black lines) were fit to the post-impact points. Between 26 and 67 days after impact, eclipses (black triangles) and occultations (red squares) were temporally separated and thus could be independently fit. All other points (blue circles) represent either single events or blended eclipses and occultations.}
    \label{fig:extinction}
\end{figure}

The model versus observed event depths were used to characterize the timescale on which the ejecta dissipated (Figure \ref{fig:extinction}). Prior to impact, variations in $\Delta M$ provided a quantitative assessment of how well the model represented the data. For both primary and secondary events the standard deviation for $\Delta M$ was $\sim0.01$ mag, suggesting that this is the noise floor below which we were unable to resolve meaningful signatures due to combined uncertainties in the data, the lightcurve decomposition process, and the mutual event models. This floor is consistent with our data quality requirements for accepted lightcurves (\S\ref{sec:analysis}).

After impact, there was a clear signature of positive $\Delta M$ that then gradually decays. We treated the $\Delta M$ values for primary and secondary events separately, and independently fit them with exponential decay curves. Both decay curves show initial values around 0.04 magnitudes. Of course this ``initial" value is tied to the first post-impact mutual event which wasn't detected until 29 hours after impact. Lightcurves collected $<29$ h after impact were not decomposable and thus not considered here. 

Surprisingly the primary and secondary events displayed different $\Delta M$ decay profiles. We found that the $\Delta M$ half life for secondary events was 24.2 $\pm$ 4.8 d, fully consistent with the bulk photometric fading of 23.7 d presented in \citet{Graykowski23}. However, the corresponding timescale for primary events was longer at 48.6 $\pm$ 12.2 d. The primary decay curve also did not return to zero at the end of the apparition, but instead remained slightly offset by $\sim0.01$ magnitude. It seems implausible that ejecta would still be affecting event depths 5 months after impact. The reasons for these issues are not obvious, but we do discuss possibilities below that motivate avenues for future work. Regardless of the cause(s) of these offsets in event depth, the close correspondence between data and model in event timing and general morphology suggests that macroscopic ejecta like the boulders seen in Hubble Space Telescope images \citep{Jewitt23} or optically thick clouds did not contribute significantly to the mutual event signatures.

%%%%%%%%%%%%%%%%
% MAGS
%%%%%%%%%%%%%%%%

\section{Photometric fading of the system} \label{sec:mags}

\begin{figure}[h!]
    \centering
    \includegraphics[width=\textwidth]{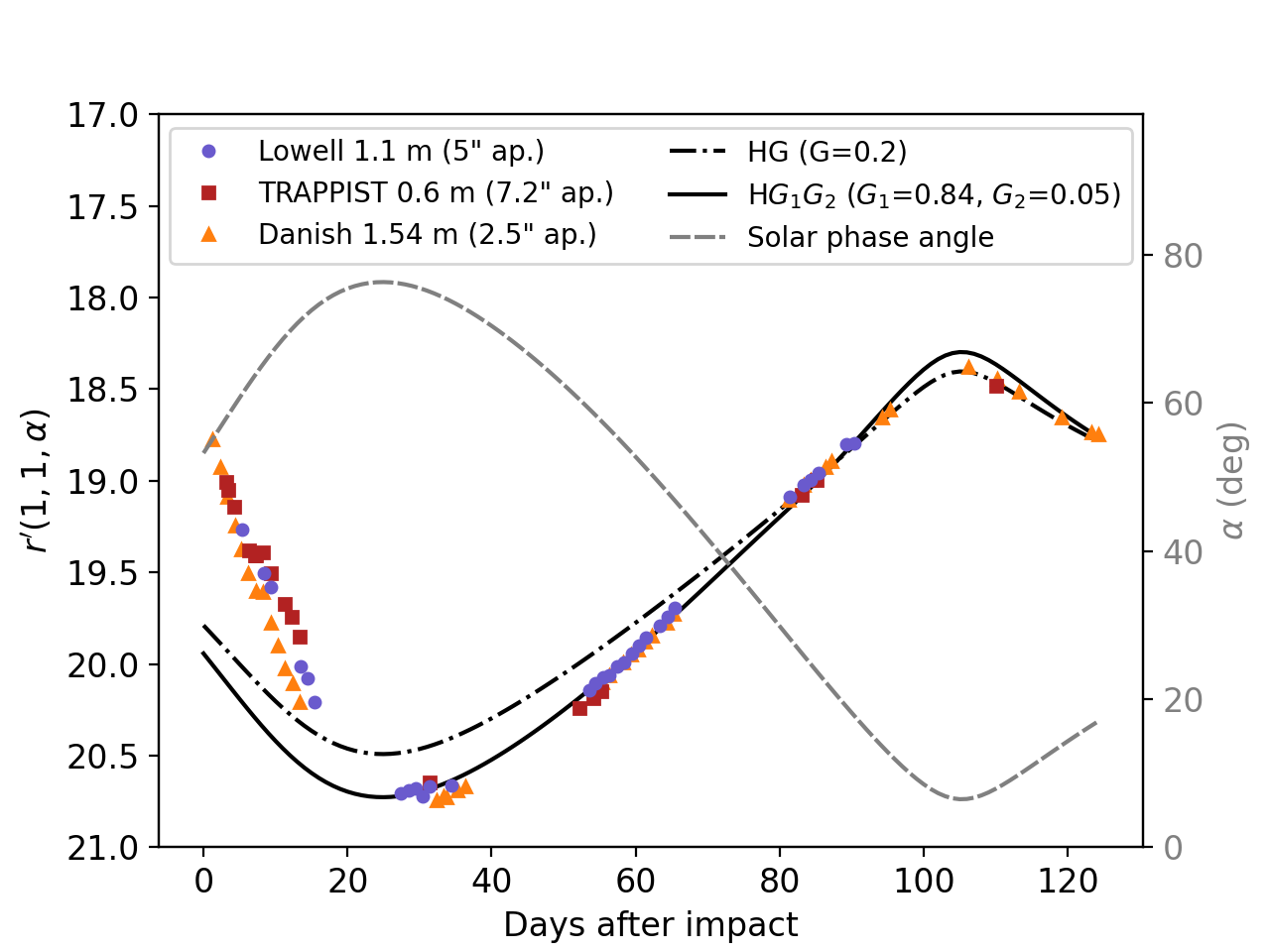}
    \caption{Apparent $r$-band magnitudes, normalized to geo- and heliocentric ranges of 1 au, of the Didymos system after DART impact. These curves include both phase angle affects and fading of ejecta. All magnitudes have been converted to the PanSTARRS $r$ filter (see text for details). Three data sets are shown: Lowell 1.1 m (blue circles), TRAPPIST 0.6 m (red squares), and the Danish 1.54 m (orange triangles). The photometric aperture for each data set is given in the legend. Both HG \citep{Bowell89} and HG$_1$G$_2$ \citep{Muinonen10} photometric models are plotted in black. The solar phase angle $\alpha$ (dashed gray curve) spans a wide range of values from a maximum of $76^\circ$ down to a minimum of $6^\circ$.}
    \label{fig:fade}
\end{figure}

Several facilities that contributed to the lightcurve campaign collected data for 3-4 months after impact. These extended data sets provided a means to monitor fading of ejecta and the photometric phase curve of the system (Figure \ref{fig:fade}). Our aim here is to address two questions: when does ejecta no longer contribute significantly to the bulk photometry and do standard photometric models represent the photometric behavior of the system? More details on photometric modeling of ground-based and in situ data are presented in \citet{Hasselmann23b}. In this analysis we focused on three data sets: those from the Lowell Observatory 1.1 m, the Danish 1.5 m, and the dual TRAPPIST 0.6 m telescopes. Each of these data sets were independently reduced using different methods (\S\ref{sec:facilities}), but internally they adopted uniform approaches across the full post-impact window and thus represent a consistent and well-calibrated baseline for long term photometric characterization. All of the Lowell photometry were measured with a 5\arcsec~aperture, the TRAPPIST data with a 7.2\arcsec~aperture, and the Danish data with a 2.5\arcsec~aperture. For each night of observation the mean magnitude outside of mutual events and mid-JD are plotted in Figure \ref{fig:fade}. These averages compensate for lightcurve variability.

Filter transforms were required to compare these data. The Lowell observations were obtained with a broad $VR$ filter and calibrated to PanSTARRS $r$. The Danish data were obtained in Johnson-Cousins $V$ or $R$, and calibrated to Cousins $R$. The TRAPPIST data were obtained with the broad Exo filters and were calibrated to Cousins $R$. All data were converted to the PanSTARRS $r$ bandpass. These conversions were based on the measured $B-V=0.795$ and $V-R=0.458$ colors of Didymos from \citet{Kitazato04}. The former was used to compute $V-r=0.22$ based on the transforms of \citet{Jester05}, which then gives $R-r=-0.24$. These colors were also used to estimate the absolute magnitude for Didymos. The mission-adopted V-band absolute magnitude $H_V=18.16$ \citep{Pravec12} converts to $H_r=17.94$, which we use below to compute the photometric phase function.

A few non-uniform aspects of these data are noted. The photometric aperture for each data set was different. This produced a clear trend in brightness as a function of aperture size, which was also discussed in \citet{Kareta23}. The data with the smallest aperture (from the Danish) appear systematically fainter, the data with the largest aperture (from TRAPPIST) were systematically brighter. However these are relatively small effects, most pronounced in the first 20 days after impact when significant ejecta was still present, and do not significantly influence the broad conclusions made here.

We have also added three nights of data, UT October 10, 11, and 12 (impact +13 to +15 d), from the Lowell 1.1 m that were not included in the lightcurve analysis (\S\ref{sec:analysis}). These data were not viable for lightcurve decomposition, but the mean magnitudes they provide are consistent with general trends and provide temporal sampling on days without other data.

The curves in Figure \ref{fig:fade} correspond to apparent magnitudes normalized to geocentric and heliocentric ranges of 1 au, and thus still include the effects of changing solar phase angle and the influence of ejecta on the total brightness of the system. This facilitates comparison to photometric phase curves computed in the convention of the IAU HG system \citep{Bowell89} and the HG$_1$G$_2$ system \citep{Muinonen10}. The value $G=0.2$ from \citet{Kitazato04} is used for the HG calculation. Values of $G_1=0.84$ and $G_2=0.05$, derived from fits to LICIACube, DRACO and ground based data \citep{Hasselmann23}, were used for the HG$_1$G$_2$ model. These particular $G_1$ and $G_2$ values are unusual, but not unprecedented, for S-type asteroids \citep{Penttila16,Mahlke21}.

The comparison of our data to these phase curves highlights several key results. First, all three data sets confirm an inflection in fading rates starting about 8 days after impact \citep{Kareta23}. This ``8 day bump" is not a result of changing viewing geometry, but may be due to increased ejecta following secondary impacts in the system. Second, the HG model does not represent the data well at phase angles $>30^\circ$, suggesting that estimates of the system brightness relative to HG predictions would be off by up $\sim0.2$ mag in the two months following impact. The predicted HG magnitudes do represent the data well later in the apparition (lunations L3 and L4) when ejecta no longer dominated the photometric signal and phase angles were lower. Finally, we estimate the system returned to pre-impact brightness about 20 days after impact. Analysis of other data sets taken in the 16-26 day window after impact, e.g. like those presented in \citet{Graykowski23} and \citet{Kareta23}, provide additional insights into when the ejecta no longer contributed a detectable enhancement. For the data presented here, interpolation from 1-15 days shows a return to the HG$_1$G$_2$ model around 20 days post-impact.

%%%%%%%%%%%%%%%%
% TAIL
%%%%%%%%%%%%%%%%

\section{Tail persistence} \label{sec:tail}

All lightcurves presented here (and even those from March 2023 that could not be used for decomposition analysis) were collected and decomposed with a tail still present. Even though the system returned to pre-impact brightness within about 20 days, the tail persisted for the entirety of the apparition with gradually decreasing brightness. While the lightcurves clearly measured the mutual events with sufficient precision to conduct the analyses described elsewhere in this paper, the interpretation of our results still depends on the influence that the tail and other ejecta played in the measured photometry. 

\begin{figure}[h!]
    \centering
    \includegraphics[width=\textwidth]{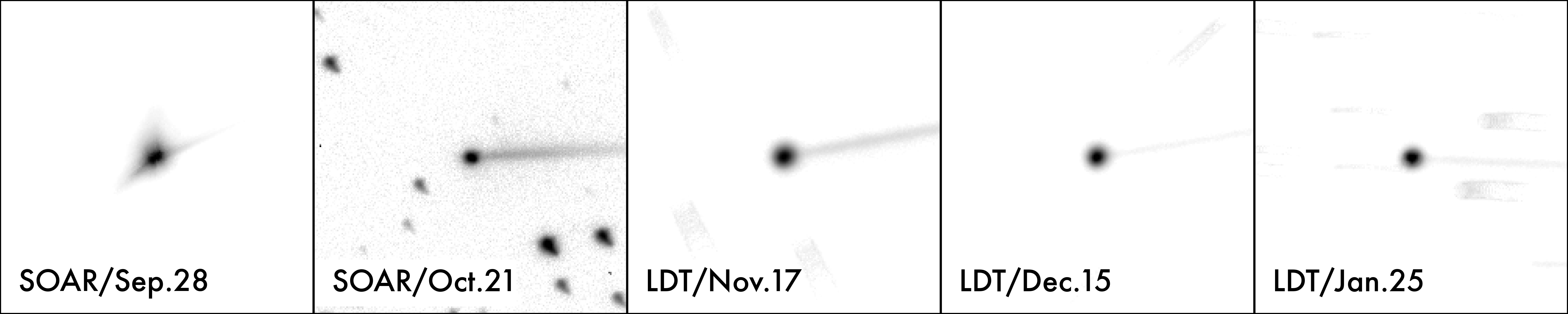}
    \caption{Comparison of the relative brightness of the tail in five stacked images from SOAR and the LDT spanning from 2022 September 28th through 2023 January 25th. All images utilized the Sloan $r$ filter, go to approximately equal depths, and are scaled logarithmically between the peak brightness of the central condensation and $0.5\%$ of that value.}
    \label{fig:persistence}
\end{figure}

In Figure \ref{fig:persistence}, we showcase five identically scaled image stacks of the Didymos system in the first four months after impact. The tail was still detectable from 4m-class telescopes an appreciable fraction of a year after impact, far beyond expectations, but it also dimmed significantly over that time period. As mentioned in Section \ref{sec:mags}, the system returned to its pre-impact brightness within several weeks of impact, suggesting that the tail (and ejecta broadly) contributed negligible brightness after that period \textit{despite} its continuing detectability. The persistence of the primary tail over such a long period implies ongoing escape of ejecta for many months after impact, and its gradual decline in brightness (optical thickness) suggests that mass loss at increasingly small rates continued well after the tail was no longer contributing significant photometric signal.

To estimate this small contribution, and to compare it against other quantities like the mutual event depths, we follow a procedure outlined in \citet{2019ApJ...881L...6S}. In essence, the brightness profile of the tail as a function of distance from the optocenter is fit well away from the central PSF and then extrapolated inwards towards the optocenter. This results in an estimate of the fraction of the brightness in the central few pixels which may be attributed to the tail. There are several caveats to this methodology that are worth noting, primarily driven by assuming that the structure of the tail thousands of kilometers from the asteroids can be extrapolated meaningfully inwards. For example, dust passing behind the asteroids may not have been fully illuminated, or dynamical effects near the binary system may have caused the dust to not be uniformly distributed. Utilizing the images in Figure \ref{fig:persistence}, the estimated contribution was as much as $\sim10\pm1\%$ of the nuclear signal on 2022 September 28, but drops to $\sim1-3\%$ on October 21st, with lower values thereafter. In other words, by the time the system returned to its bulk brightness, the contribution of the tail was similar to or smaller than typical observational errors. Furthermore, in the first few weeks after impact, where the brightness of the bulk ejecta was important, the tail was but a minor contributor. In short, the brightness of the tail played little role in the photometric conclusions drawn here, but the persistence of the tail was a clear sign of ongoing mass loss from and evolution of the system more broadly.

%%%%%%%%%%%%%%%%
% DISCUSSION
%%%%%%%%%%%%%%%%

\section{Results \& Discussion} \label{sec:disc}

We have presented an overview of the lightcurve photometry campaign carried out in support of the DART mission. Data and analysis were presented from the impact apparition, spanning roughly 8 months from July 2022 to February 2023. This large data set included 224 lightcurves from 28 telescopes, over 38,000 individual exposures, and represented over 1000 hours spent targeting the Didymos system (Table \ref{tab:observatories}). The duration and observing circumstances (e.g. galactic latitude, declination, apparent magnitude) of this campaign necessitated a coordinated plan that leveraged telescopes across a wide ranges of aperture sizes and geographic locations (Figures \ref{fig:obs}, \ref{fig:map}). A series of 10 observing windows (lunations) were defined during the impact apparition (Table \ref{tab:lunations}) to maximize data quality and to establish times when data collection rates were expected to be highest, thus providing the best opportunity for yielding sets of decomposable lightcurves. A set of practice targets were defined prior to the start of the apparition as a way for observers to establish observing and reduction protocols on targets with similar observing circumstances (e.g. apparent magnitude, non-sidereal rates, lunar phase) to Didymos in the weeks following impact. These preparatory steps were important to achieve strict signal-to-noise requirements; the mean rms residual across the full data set was 0.0073 mag, with some individual lightcurves showing residuals better than 0.004 mag.

 Each observer employed their own data reduction methods. This was an intentional part of the campaign to avoid introducing systematic errors that might arise from a single, uniform approach to data reduction. Given the volume of data, subtle differences in reduction methods were unlikely to have much influence on the end results. However, one consequence of this approach was that the error bars on the reported photometry were not mutually consistent. As such we ignored the reported error bars in the decomposition analysis (Section \ref{sec:analysis}) and assessed the quality of each lightcurve based on rms residuals relative to the best fit Fourier models.

The full suite of lightcurves was divided into 43 decomposable sets (Table \ref{tab:decomp}), where a set was defined based on constant morphology of the primary rotational signature from Didymos. For a given set the decomposition process included light time and geometric corrections, masking of data taken during mutual events, and then fitting data outside of the mask with a n-th order Fourier series to isolate dips in brightness associated with mutual events. Included in this data set is the first post-impact mutual event captured just 29 hours after DART impacted Dimorphos (Figure \ref{fig:sep28}), as well as simultaneous coverage of some events by as many as four independent observatories. Consistency across these multiply covered events provides an important validation of data reduction methods (Figure \ref{fig:resid}). The decomposition of lightcurves into their constituent parts -- primary rotation and mutual events -- provided the foundation for modeling the rotational and orbital dynamics of the Didymos system \citep{Naidu23,Scheirich23}. To encourage future work with these data, we have included here a supplementary file with the full set of lightcurve measurements and decomposed residuals.

We have leveraged select aspects of this data set to investigate the post-impact photometric evolution of the system. While the focus on mutual events has been to derive their timing to serve as a constraint on the orbital period of Dimorphos \citep{Naidu23,Scheirich23}, the depths of events also contain information about the influence of ejecta on the measured photometry. To facilitate this analysis we compare the measured depths of mutual events to model predictions by \citet{Naidu23}. This comparison shows that prior to impact the model reasonably represents the timing, depth, and gross morphology of mutual events (Figure \ref{fig:models}). However, after impact the measured event depths are systematically shallower than predicted by models (Figure \ref{fig:extinction}). This is a clear indication of additional flux from residual ejecta acting to dampen the event depths in the observations. An initial dampening of $\sim0.04$ magnitudes for the first detected mutual event on 29 September 2022 shows that the ejecta cloud is already optically thin at that time. We also fit decay curves to the post-impact offsets in mutual event depths ($\Delta M$ = model - observation). Secondary eclipses and occultations show behavior consistent with other constraints \citep[e.g.][]{Graykowski23}, namely the mutual event depths return to model predictions with a half life of $24.2 \pm 4.8$ days. However, interpreting the analysis of primary events is not as straightforward. Our analysis shows that the measured primary events do not return to the model at the end of the apparition. Furthermore, the decay timescale for primary events is longer at $48.6 \pm 12.2$ days. These different outcomes for primary (Didymos) versus secondary (Dimorphos) mutual events may be due to a number of factors.

The apparent differences in the $\Delta M$ curves could be simply attributed to uncertainties in the data relative to the parameters being fit. The reported uncertainties on the decay timescales are $1\sigma$, thus the two are distinct at only the $1.4\sigma$ level. Also possible is that the assumption of exponential decay may not be a good representation for the complex evolution of ejecta in the system. That certainly appears to be the case for primary events where $\Delta M$ is roughly constant across the first two post-impact lunations before dropping off in the 3rd lunation, $\sim50$ days after impact (Figure \ref{fig:extinction}).

Assumptions built into the mutual event model could also be affecting the $\Delta M$ decay curves. Incorrect assumptions about the body shapes and/or surface properties like albedo and roughness, would lead to discrepancies between the model and observations. The biggest offsets between data and model are seen for primary events in the L1 lunation (Figure \ref{fig:extinction}). These occur when the system was at its maximum solar phase angle  $\alpha>70^\circ$. It is possible that assumptions about photometric scattering and projected shapes break down at such viewing geometries. However, it is unclear how such assumptions would have an asymmetric affect, e.g. different decay timescales in primary versus secondary events. 

The $\Delta M$ discrepancy could also be attributed to the fact that secondary events are easier to reproduce with these models because they are less sensitive to the detailed properties of Dimorphos. In other words, when Dimorphos is in total eclipse or occultation, it no longer contributes to the measured flux of the system and thus its detailed properties don't influence predicted depths for secondary mutual events. However, during primary mutual events the detailed properties of Dimorphos could have a more significant influence based on the specific regions on the surface of Didymos that are being shadowed or blocked. This conceptual interpretation seems reasonable and suggests that the model of \citet{Naidu23} more accurately predicted the photometric behavior of secondary mutual events.

Further work is clearly needed to investigate some of these details in the data as well as assumptions built into the mutual event model. More stringent vetting of data to consider only the highest quality detections that cover both ingress and egress at high temporal resolution could help to mitigate uncertainty and scatter in the comparisons to the model. The model itself makes assumptions about the body shapes and scattering properties. Evidence for post-impact modifications to the shape of Dimorphos \citep{Naidu23} could have a significant influence on both occultation and eclipse depths. These effects would be most pronounced at high solar phase angles (e.g. lunation L1) when occultations and eclipses were well separated in time and occurred many tens of degrees apart in the mean anomaly of Dimorphos. Higher order dynamical effects such as libration and precesion of Dimorphos \citep{Naidu23} could also be complicating the interpretation of the data. It is also possible that the assumed Lommel-Seeliger properties for S-type asteroids \citep{Huang17} adopted by the mutual event model may need to be revised.

We have also presented a simple phase curve analysis based on nightly average magnitudes from a few select observatories (Figure \ref{fig:fade}). This analysis showed that canonical HG magnitude predictions \citep{Bowell89} do not represent the data well, particularly at phase angles $>40^\circ$. An HG$_1$G$_2$ model \citep{Muinonen10} does reproduce the data reasonably well with parameters adopted from \citet{Hasselmann23}. However, the parameters used in this HG$_1$G$_2$ model are unusual for S-type asteroids \citep{Penttila16,Mahlke21}.  These issues could be related to differences in the times and viewing geometries of our data relative to those used to derive the model parameters. Disentangling the effects of shape and viewing geometry on photometric phase curves can be challenging for near-Earth asteroids \citep[e.g.][]{Jackson22}. Further work is needed to address whether discrepancies with the mutual event model (e.g. Figure \ref{fig:extinction}) could be attributed to these potentially unusual photometric properties of the Didymos system.

Lastly, we noted that all of our post-impact lightcurve photometry was performed in the presence of an ejecta cloud (for several weeks after impact) and a persistent tail that was still detectable 5 months after impact. We showed that by 1 month after impact dust in the tail contributed $<1\%$ of the nuclear signal and thus dropped below typical photometric error bars.

While the data and analysis presented here have contributed to successful completion of the DART mission's Level 1 requirements \citep{Chabot23}, open questions remain related to the photometric behavior of Didymos in the months after impact, the detailed morphology of mutual events, the role that ejecta played in the observed properties of the system, and the extent to which higher order dynamical effects are detectable in these data. Fortunately, a number of these questions will be resolved when the European Space Agency's Hera mission arrives in the Didymos system in late 2026 \citep{Michel23}. Hera, in its nominal mission, will rendezvous with Didymos and Dimorphos for a 6 month investigation of surface and interior properties, as well as studies of any residual impact ejecta that may still be present in the system. Until then ground-based observing opportunities in 2024 and 2025 will continue to provide deeper understanding of the post-DART Didymos system.

%%%%%%%%%%%%%%%%
% ACKNOWLEDGEMENTS
%%%%%%%%%%%%%%%%
%\begin{acknowledgments}

\clearpage
{\bf Acknowledgments}

This work is based upon a large volume of data collected at dozens of observatories around the world. We are deeply grateful to all of the telescope operators, engineers, instrument scientists, and site managers who enabled these observations. We acknowledge the thoughtful reviews provided by two anonymous referees.

Significant portions of this work were completed at Lowell Observatory, which sits at the base of mountains sacred to tribes throughout the region. We honor their past, present, and future generations, who have lived here for millennia and will forever call this place home.

% FUNDING SOURCES
Much of this work were supported by the DART mission under NASA Contract No. 80MSFC20D0004. NM and TK acknowledge additional funding from NASA grants 80NSSC21K1328 and NNX17AH06G, awarded in support of the Mission Accessible Near-Earth Object Survey (MANOS). The work of S.P.N. and S.R.C. was carried out at the Jet Propulsion Laboratory, California Institute of Technology, under a contract with the National Aeronautics and Space Administration (No. 80NM0018D0004). The work at Ond\v{r}ejov and observations with the Danish telescope on La Silla were supported by the Grant Agency of the Czech Republic, grants 20-04431S and 23-4946S. Access to computing and storage facilities owned by parties and projects contributing to the National Grid Infrastructure MetaCentrum provided under the program ``Projects of Large Research, Development, and Innovations Infrastructures" (CESNET LM2015042), and the CERIT Scientific Cloud LM2015085, is greatly appreciated. YK thanks support from the European Federation of Academies of Sciences and Humanities (grant ALLEA EFDS-FL1-18). IR was funded by the Aerospace Committee of the Ministry of Digital Development, Innovations and Aerospace Industry of the Republic of Kazakhstan (Grant No. BR 11265408). YK, OB, KE are grateful to the staff of the Maidanak Observatory for their support during the observations. AR and CS acknowledge support from the UK Science and Technology Facilities Council. This project has received funding from the European Union’s Horizon 2020 research and innovation programme under grant agreement No 870403 (NEOROCKS). PLP was partly funded by Programa de Iniciación en Investigación-Universidad de Antofagasta. INI-17-03. The work of MH was supported by the Slovak Grant Agency for Science VEGA (Grant No. 2/0059/22) and by the Slovak Research and Development Agency under the Contract No. APVV-19-0072. TSR acknowledges funding from Ministerio de Ciencia e Innovación (Spanish Government), PGC2021, PID2021-125883NB-C21. This work was (partially) supported by the Spanish MICIN/AEI/10.13039/501100011033 and by "ERDF A way of making Europe" by the “European Union” through grant PID2021-122842OB-C21, and the Institute of Cosmos Sciences University of Barcelona (ICCUB, Unidad de Excelencia ’María de Maeztu’) through grant CEX2019-000918-M. The work of MP and DB  was supported by a grant of the Romanian National Authority for Scientific  Research -- UEFISCDI, project number PN-III-P2-2.1-PED-2021-3625.

% TELESCOPES/INSTRUMENTS/FACILITIES
These results made use of the Lowell Discovery Telescope (LDT) at Lowell Observatory.  Lowell is a private, non-profit institution dedicated to astrophysical research and public appreciation of astronomy and operates the LDT in partnership with Boston University, the University of Maryland, the University of Toledo, Northern Arizona University and Yale University. The Large Monolithic Imager was built by Lowell Observatory using funds provided by the National Science Foundation (AST-1005313). This work is based in part on observations obtained at the Southern Astrophysical Research (SOAR) telescope, which is a joint project of the Ministério da Ciência, Tecnologia e Inovações do Brasil (MCTI/LNA), the US National Science Foundation’s NOIRLab, the University of North Carolina at Chapel Hill (UNC), and Michigan State University (MSU). This research is based on observations made with the 1.0m Jacobus Kapteyn Telescope at the Roque de los Muchachos Observatory on La Palma and the 0.8m IAC80 telescope at the Teide Observatory on Tenerife. DO was supported by grant No. 2022/B/ST9/00267 from the National Science Center, Poland. VT was supported by the National Scholarship programme of the Slovak Republic - academic year 2023/24. TRAPPIST is a project funded by the Belgian Fonds (National) de la Recherche Scientifique (F.R.S.-FNRS) under grant PDR T.0120.21. TRAPPIST-North is a project funded by the University of Liège, in collaboration with the Cadi Ayyad University of Marrakech (Morocco). E. Jehin is a F.R.S.-FNRS Senior Research Associate. This work has made use of data from the Asteroid Terrestrial-impact Last Alert System (ATLAS) project. ATLAS is primarily funded to search for near earth asteroids through NASA grants NN12AR55G, 80NSSC18K0284, and 80NSSC18K1575; byproducts of the NEO search include images and catalogs from the survey area. The ATLAS science products have been made possible through the contributions of the University of Hawaii Institute for Astronomy, the Queen’s University Belfast, the Space Telescope Science Institute, and the South African Astronomical Observatory. This paper is based on observations made with the MuSCAT3 instrument, developed by the Astrobiology Center and under financial supports by JSPS KAKENHI (JP18H05439) and JST PRESTO (JPMJPR1775), at Faulkes Telescope North on Maui, HI, operated by the Las Cumbres Observatory. This paper was partially based on observations obtained at the Bohyunsan Optical Astronomy Observatory (BOAO), which is operated by the Korea Astronomy and Space Science Institute (KASI).

%\end{acknowledgments}


\vspace{5mm}
\facilities{Magellan:Baade (IMACS), LDT (LMI), SOAR (Goodman), FTN (MuSCAT3), VATT, BOAO:1.8m, Danish 1.54m Telescope (DFOSC), Maidanak:1.5m, Sanchez (MuSCAT2), Hall (NASA42), LCOGT (Sinistro), ING:Kapteyn, Swope, Spacewatch:0.9m, OT:0.8m (CAMELOT2), OO:0.65, TRAPPIST.}


\software{astropy \citep{astropy13,astropy18},  
          Source Extractor \citep{Bertin96},
          Scamp \citep{Bertin06},
          sbpy \citep{Mommert19},
          Photometry Pipeline \citep{Mommert17}
          Tycho Tracker \citep{Parrott20}
          astroquery \citep{Ginsburg19}
          }



\appendix
\restartappendixnumbering

\section{Observational Details}

\startlongtable
\begin{deluxetable*}{lllccr}
\tablecaption{Observational details for each lightcurve presented in this work. Line breaks represent independent decompositions, as indicated by the Decomposition ID. \label{tab:details}}
\tabletypesize{\scriptsize}
\tablehead{\colhead{Facility} & \colhead{UTC Start} & \colhead{JD Range} & \colhead{Duration (h)} & \colhead{Data Points} & \colhead{Decomposition ID}}
\startdata
6.5 m Magellan & 2022-07-02T04:00 & 2459762.66699-2459762.94198 & 6.6 & 193 & pre-L1\\ 
4.1 m SOAR & 2022-07-04T06:52 & 2459764.78643-2459764.94339 & 3.8 & 129 & pre-L1\\ 
4.1 m SOAR & 2022-07-05T04:24 & 2459765.68348-2459765.94987 & 6.4 & 210 & pre-L1\\ 
4.3 m LDT & 2022-07-06T08:02 & 2459766.83477-2459766.96810 & 3.2 & 89 & pre-L1\\ 
4.3 m LDT & 2022-07-07T07:50 & 2459767.82686-2459767.97285 & 3.5 & 85 & pre-L1\\ 
\hline
1 m LCOGT & 2022-07-31T03:36 & 2459791.65044-2459791.90579 & 6.1 & 125 & pre-L2\\ 
\hline
1 m JKT & 2022-08-18T00:45 & 2459809.53134-2459809.71404 & 4.4 & 119 & pre-L3.1\\ 
1 m JKT & 2022-08-19T00:49 & 2459810.53460-2459810.71368 & 4.3 & 110 & pre-L3.1\\ 
\hline
1 m Swope & 2022-08-22T01:37 & 2459813.56770-2459813.91689 & 8.4 & 498 & pre-L3.2\\ 
1 m Swope & 2022-08-23T01:25 & 2459814.55946-2459814.92080 & 8.7 & 517 & pre-L3.2\\ 
0.7 m AC-32 & 2022-08-23T20:32 & 2459815.35618-2459815.55030 & 4.7 & 88 & pre-L3.2\\ 
1 m Swope & 2022-08-24T01:29 & 2459815.56200-2459815.92027 & 8.6 & 518 & pre-L3.2\\ 
0.7 m AC-32 & 2022-08-24T21:09 & 2459816.38131-2459816.54553 & 3.9 & 82 & pre-L3.2\\ 
1 m Swope & 2022-08-25T03:27 & 2459816.64420-2459816.91983 & 6.6 & 397 & pre-L3.2\\ 
0.7 m AC-32 & 2022-08-25T22:55 & 2459817.45536-2459817.54141 & 2.1 & 36 & pre-L3.2\\ 
\hline
1 m Swope & 2022-08-30T00:51 & 2459821.53574-2459821.91384 & 9.1 & 531 & pre-L3.3\\ 
1 m Swope & 2022-08-31T00:53 & 2459822.53737-2459822.91240 & 9.0 & 534 & pre-L3.3\\ 
1 m Swope & 2022-09-01T01:00 & 2459823.54167-2459823.91311 & 8.9 & 523 & pre-L3.3\\ 
1 m Swope & 2022-09-02T00:49 & 2459824.53435-2459824.91263 & 9.1 & 550 & pre-L3.3\\ 
\hline
0.6 m TS & 2022-09-04T02:23 & 2459826.59957-2459826.86723 & 6.4 & 153 & pre-L3.4\\ 
0.6 m TS & 2022-09-05T02:22 & 2459827.59877-2459827.69795 & 2.4 & 81 & pre-L3.4\\ 
0.6 m TS & 2022-09-06T02:33 & 2459828.60646-2459828.67713 & 1.7 & 52 & pre-L3.4\\ 
\hline
0.6 m TS & 2022-09-12T01:31 & 2459834.56367-2459834.74939 & 4.5 & 256 & pre-L4.1\\ 
1 m LCOGT & 2022-09-12T14:10 & 2459835.09086-2459835.27470 & 4.4 & 198 & pre-L4.1\\ 
1 m LCOGT & 2022-09-12T20:01 & 2459835.33419-2459835.51788 & 4.4 & 187 & pre-L4.1\\ 
0.6 m TS & 2022-09-14T06:33 & 2459836.77356-2459836.90962 & 3.3 & 147 & pre-L4.1\\ 
0.6 m TS & 2022-09-15T04:34 & 2459837.69065-2459837.81569 & 3.0 & 145 & pre-L4.1\\ 
1.5 m Danish & 2022-09-16T08:22 & 2459838.84882-2459838.91024 & 1.5 & 98 & pre-L4.1\\ 
\hline
0.6 m TS & 2022-09-18T07:49 & 2459840.82589-2459840.90635 & 1.9 & 76 & pre-L4.2\\ 
0.5 m T72 & 2022-09-20T01:44 & 2459842.57286-2459842.74630 & 4.2 & 162 & pre-L4.2\\ 
1.5 m Danish & 2022-09-20T02:38 & 2459842.61009-2459842.90694 & 7.1 & 463 & pre-L4.2\\ 
0.6 m TS & 2022-09-20T02:46 & 2459842.61532-2459842.70663 & 2.2 & 120 & pre-L4.2\\ 
0.6 m TS & 2022-09-20T07:16 & 2459842.80324-2459842.90456 & 2.4 & 139 & pre-L4.2\\ 
0.6 m TS & 2022-09-21T03:46 & 2459843.65757-2459843.90548 & 5.9 & 243 & pre-L4.2\\ 
0.6 m TS & 2022-09-22T08:53 & 2459844.87053-2459844.90641 & 0.9 & 45 & pre-L4.2\\ 
\hline
0.6 m TS & 2022-09-23T06:47 & 2459845.78307-2459845.90582 & 2.9 & 161 & pre-L4.3\\ 
0.6 m TS & 2022-09-24T02:03 & 2459846.58601-2459846.89954 & 7.5 & 264 & pre-L4.3\\ 
0.6 m TS & 2022-09-25T02:11 & 2459847.59151-2459847.89606 & 7.3 & 375 & pre-L4.3\\ 
1 m Swope & 2022-09-25T03:16 & 2459847.63677-2459847.89319 & 6.2 & 435 & pre-L4.3\\ 
1.5 m Danish & 2022-09-25T03:28 & 2459847.64511-2459847.89944 & 6.1 & 569 & pre-L4.3\\ 
1 m LCOGT & 2022-09-25T14:50 & 2459848.11830-2459848.24055 & 2.9 & 172 & pre-L4.3\\ 
1.5 m Danish & 2022-09-26T03:50 & 2459848.66035-2459848.70365 & 1.0 & 92 & pre-L4.3\\ 
1 m Swope & 2022-09-26T02:15 & 2459848.59377-2459848.88760 & 7.1 & 429 & pre-L4.3\\ 
0.6 m TS & 2022-09-26T07:41 & 2459848.82069-2459848.89597 & 1.8 & 113 & pre-L4.3\\ 
\hline
1 m Swope & 2022-09-28T02:33 & 2459850.60635-2459850.75643 & 3.6 & 237 & L0.1\\ 
1.5 m Danish & 2022-09-28T02:38 & 2459850.60989-2459850.88674 & 6.6 & 340 & L0.1\\ 
\hline
1 m Swope & 2022-09-29T02:40 & 2459851.61163-2459851.88686 & 6.6 & 433 & L0.2\\ 
1.5 m Danish & 2022-09-29T02:50 & 2459851.61825-2459851.89795 & 6.7 & 639 & L0.2\\ 
1 m LCOGT & 2022-09-29T04:52 & 2459851.70320-2459851.86630 & 3.9 & 212 & L0.2\\ 
\hline
0.6 m TN & 2022-09-30T02:09 & 2459852.58974-2459852.72890 & 3.3 & 94 & L0.3\\ 
1.5 m Danish & 2022-09-30T02:40 & 2459852.61163-2459852.90395 & 7.0 & 669 & L0.3\\ 
1 m Swope & 2022-09-30T02:52 & 2459852.61965-2459852.88542 & 6.4 & 420 & L0.3\\ 
1.5 m TCS & 2022-09-30T03:36 & 2459852.65033-2459852.73971 & 2.1 & 86 & L0.3\\ 
1 m LCOGT & 2022-09-30T03:52 & 2459852.66165-2459852.88666 & 5.4 & 319 & L0.3\\ 
0.6 m TS & 2022-09-30T07:03 & 2459852.79385-2459852.88887 & 2.3 & 87 & L0.3\\ 
\hline
1 m LCOGT & 2022-09-30T21:45 & 2459853.40690-2459853.55059 & 3.4 & 168 & L0.4\\ 
1 m Swope & 2022-10-01T03:28 & 2459853.64455-2459853.88504 & 5.8 & 376 & L0.4\\ 
0.6 m TS & 2022-10-01T03:29 & 2459853.64552-2459853.88815 & 5.8 & 246 & L0.4\\ 
1 m LCOGT & 2022-10-01T04:00 & 2459853.66732-2459853.88294 & 5.2 & 292 & L0.4\\ 
1.5 m Danish & 2022-10-01T06:33 & 2459853.77308-2459853.89512 & 2.9 & 278 & L0.4\\ 
\hline
1 m LCOGT & 2022-10-01T21:48 & 2459854.40896-2459854.63187 & 5.3 & 268 & L0.5\\ 
1.5 m Danish & 2022-10-02T03:05 & 2459854.62906-2459854.85671 & 5.5 & 530 & L0.5\\ 
1 m Swope & 2022-10-02T03:15 & 2459854.63565-2459854.88820 & 6.1 & 396 & L0.5\\ 
1 m LCOGT & 2022-10-02T04:00 & 2459854.66735-2459854.88319 & 5.2 & 269 & L0.5\\ 
1.1 m Hall & 2022-10-02T08:19 & 2459854.84674-2459854.95490 & 2.6 & 132 & L0.5\\ 
\hline
1 m LCOGT & 2022-10-02T22:00 & 2459855.41728-2459855.53916 & 2.9 & 136 & L0.6\\ 
1 m LCOGT & 2022-10-03T01:00 & 2459855.54228-2459855.62661 & 2.0 & 99 & L0.6\\ 
1 m Swope & 2022-10-03T03:29 & 2459855.64571-2459855.89431 & 6.0 & 385 & L0.6\\ 
0.6 m TS & 2022-10-03T03:33 & 2459855.64845-2459855.87218 & 5.4 & 275 & L0.6\\ 
1 m LCOGT & 2022-10-03T04:05 & 2459855.67017-2459855.86302 & 4.6 & 248 & L0.6\\ 
1.5 m Danish & 2022-10-03T04:27 & 2459855.68573-2459855.72804 & 1.0 & 98 & L0.6\\ 
\hline
0.6 m TN & 2022-10-04T02:13 & 2459856.59250-2459856.70134 & 2.6 & 93 & L0.7\\ 
0.6 m TS & 2022-10-04T03:43 & 2459856.65517-2459856.86429 & 5.0 & 92 & L0.7\\ 
1 m Swope & 2022-10-04T03:46 & 2459856.65705-2459856.85069 & 4.6 & 224 & L0.7\\ 
1 m LCOGT & 2022-10-04T04:15 & 2459856.67770-2459856.86631 & 4.5 & 248 & L0.7\\ 
\hline
1 m LCOGT & 2022-10-04T23:15 & 2459857.46932-2459857.62841 & 3.8 & 206 & L0.8\\ 
1.5 m Danish & 2022-10-05T03:40 & 2459857.65293-2459857.74436 & 2.2 & 151 & L0.8\\ 
0.6 m TS & 2022-10-05T03:47 & 2459857.65778-2459857.74651 & 2.1 & 94 & L0.8\\ 
1.1 m Hall & 2022-10-05T08:03 & 2459857.83609-2459858.00200 & 4.0 & 142 & L0.8\\ 
1 m Swope & 2022-10-05T03:45 & 2459857.65664-2459857.77734 & 2.9 & 181 & L0.8\\
\hline
1 m LCOGT & 2022-10-05T22:18 & 2459858.42973-2459858.58280 & 3.7 & 194 & L0.9\\ 
1 m Swope & 2022-10-06T03:49 & 2459858.65972-2459858.88938 & 5.5 & 346 & L0.9\\ 
1 m LCOGT & 2022-10-06T04:31 & 2459858.68822-2459858.83991 & 3.6 & 194 & L0.9\\ 
1.5 m Danish & 2022-10-06T05:03 & 2459858.71052-2459858.89416 & 4.4 & 370 & L0.9\\ 
0.6 m TS & 2022-10-06T05:16 & 2459858.71951-2459858.85093 & 3.2 & 110 & L0.9\\ 
\hline
1.5 m Danish & 2022-10-07T04:10 & 2459859.67378-2459859.85302 & 4.3 & 379 & L0.10\\ 
1.5 m TCS & 2022-10-07T02:49 & 2459859.61755-2459859.75465 & 3.3 & 114 & L0.10\\ 
1 m LCOGT & 2022-10-07T04:52 & 2459859.70316-2459859.88078 & 4.3 & 219 & L0.10\\ 
\hline
1 m LCOGT & 2022-10-07T22:31 & 2459860.43826-2459860.62727 & 4.5 & 246 & L0.11\\ 
0.6 m TS & 2022-10-08T03:52 & 2459860.66117-2459860.73576 & 1.8 & 88 & L0.11\\ 
1.5 m Danish & 2022-10-08T04:00 & 2459860.66732-2459860.89540 & 5.5 & 471 & L0.11\\ 
\hline
1 m LCOGT & 2022-10-08T22:37 & 2459861.44249-2459861.62691 & 4.4 & 200 & L0.12\\ 
0.6 m TS & 2022-10-09T04:08 & 2459861.67266-2459861.86428 & 4.6 & 165 & L0.12\\ 
1 m LCOGT & 2022-10-09T04:49 & 2459861.70073-2459861.88372 & 4.4 & 244 & L0.12\\ 
1.5 m Danish & 2022-10-09T06:41 & 2459861.77911-2459861.89363 & 2.7 & 244 & L0.12\\ 
\hline
0.6 m TS & 2022-10-10T04:13 & 2459862.67614-2459862.86424 & 4.5 & 246 & L0.13\\ 
1.5 m Danish & 2022-10-10T04:30 & 2459862.68801-2459862.88784 & 4.8 & 419 & L0.13\\ 
1 m LCOGT & 2022-10-10T04:49 & 2459862.70070-2459862.79332 & 2.2 & 102 & L0.13\\ 
\hline
1.5 m TCS & 2022-10-17T02:50 & 2459869.61814-2459869.75936 & 3.4 & 114 & L1.1\\ 
0.9 m Spacewatch & 2022-10-17T10:43 & 2459869.94658-2459870.02078 & 1.8 & 51 & L1.1\\ 
\hline
0.9 m Spacewatch & 2022-10-20T09:01 & 2459872.87591-2459873.00474 & 3.1 & 92 & L1.2\\ 
2.4 m MRO & 2022-10-21T08:27 & 2459873.85238-2459873.99883 & 3.5 & 202 & L1.2\\ 
0.9 m Spacewatch & 2022-10-21T08:40 & 2459873.86150-2459874.01733 & 3.7 & 81 & L1.2\\ 
\hline
1.1 m Hall & 2022-10-24T08:06 & 2459876.83777-2459877.00597 & 4.0 & 81 & L1.3\\ 
1.1 m Hall & 2022-10-25T08:04 & 2459877.83659-2459878.00060 & 3.9 & 88 & L1.3\\ 
0.9 m Spacewatch & 2022-10-25T09:36 & 2459877.90001-2459878.01528 & 2.8 & 72 & L1.3\\ 
1.1 m Hall & 2022-10-26T08:05 & 2459878.83737-2459878.99494 & 3.8 & 91 & L1.3\\ 
1.8 m VATT & 2022-10-26T09:01 & 2459878.87629-2459879.01152 & 3.2 & 114 & L1.3\\ 
0.9 m Spacewatch & 2022-10-26T09:43 & 2459878.90541-2459879.01919 & 2.7 & 57 & L1.3\\
1.1 m Hall & 2022-10-27T08:00 & 2459879.83398-2459880.00799 & 4.2 & 94 & L1.3\\ 
0.9 m Spacewatch & 2022-10-27T10:20 & 2459879.93084-2459880.02873 & 2.3 & 69 & L1.3\\ 
\hline
0.6 m Ondrejov & 2022-10-28T00:59 & 2459880.54132-2459880.68111 & 3.4 & 94 & L1.4\\ 
1 m LCOGT & 2022-10-28T02:21 & 2459880.59792-2459880.75474 & 3.8 & 162 & L1.4\\ 
0.6 m TS & 2022-10-28T05:45 & 2459880.73984-2459880.87327 & 3.2 & 56 & L1.4\\ 
1.5 m Danish & 2022-10-29T07:18 & 2459881.80431-2459881.87030 & 1.6 & 69 & L1.4\\ 
1.1 m Hall & 2022-10-28T07:58 & 2459880.83197-2459881.01225 & 4.3 & 120 & L1.4\\ 
1.1 m Hall & 2022-10-29T07:50 & 2459881.82682-2459882.01607 & 4.5 & 110 & L1.4\\ 
1.8 m VATT & 2022-10-29T08:49 & 2459881.86752-2459882.02402 & 3.8 & 100 & L1.4\\ 
1 m LCOGT & 2022-10-30T02:16 & 2459882.59474-2459882.75793 & 3.9 & 178 & L1.4\\ 
0.6 m Ondrejov & 2022-10-30T02:20 & 2459882.59771-2459882.68875 & 2.2 & 36 & L1.4\\ 
1.5 m Danish & 2022-10-30T05:16 & 2459882.71997-2459882.86674 & 3.5 & 151 & L1.4\\ 
1.1 m Hall & 2022-10-31T07:47 & 2459883.82451-2459883.99910 & 4.2 & 124 & L1.4\\ 
1.5 m Danish & 2022-11-01T05:24 & 2459884.72555-2459884.86775 & 3.4 & 94 & L1.4\\ 
1.5 m Danish & 2022-11-02T05:27 & 2459885.72764-2459885.86534 & 3.3 & 92 & L1.4\\ 
0.9 m Spacewatch & 2022-11-02T09:27 & 2459885.89397-2459886.01292 & 2.9 & 69 & L1.4\\ 
\hline
1.5 m TCS & 2022-11-17T04:56 & 2459900.70591-2459900.77040 & 1.5 & 37 & L2.1\\ 
4.3 m LDT & 2022-11-17T07:24 & 2459900.80843-2459901.05212 & 5.8 & 379 & L2.1\\ 
0.6 m TN & 2022-11-18T01:04 & 2459901.54456-2459901.75044 & 4.9 & 125 & L2.1\\ 
1 m LCOGT & 2022-11-18T01:31 & 2459901.56323-2459901.75770 & 4.7 & 174 & L2.1\\ 
\hline
4.3 m LDT & 2022-11-18T07:26 & 2459901.81037-2459902.05936 & 6.0 & 479 & L2.2\\ 
0.9 m Spacewatch & 2022-11-19T08:34 & 2459902.85749-2459903.03832 & 4.3 & 113 & L2.2\\ 
1.1 m Hall & 2022-11-19T09:18 & 2459902.88771-2459903.02975 & 3.4 & 52 & L2.2\\ 
0.6 m TN & 2022-11-20T01:04 & 2459903.54492-2459903.74612 & 4.8 & 87 & L2.2\\ 
1.1 m Hall & 2022-11-20T06:47 & 2459903.78303-2459904.04012 & 6.2 & 108 & L2.2\\ 
0.9 m Spacewatch & 2022-11-20T07:41 & 2459903.82040-2459904.03469 & 5.1 & 147 & L2.2\\ 
\hline
0.6 m TN & 2022-11-21T01:48 & 2459904.57533-2459904.74878 & 4.2 & 103 & L2.3\\ 
1.5 m Danish & 2022-11-21T05:16 & 2459904.71981-2459904.86194 & 3.4 & 77 & L2.3\\ 
1.1 m Hall & 2022-11-21T06:42 & 2459904.77930-2459905.03178 & 6.1 & 104 & L2.3\\ 
0.9 m Spacewatch & 2022-11-21T09:03 & 2459904.87735-2459905.02906 & 3.6 & 103 & L2.3\\ 
1.5 m Danish & 2022-11-22T05:15 & 2459905.71932-2459905.85242 & 3.2 & 86 & L2.3\\ 
1 m LCOGT & 2022-11-22T01:16 & 2459905.55338-2459905.76103 & 5.0 & 159 & L2.3\\ 
2.4 m MRO & 2022-11-22T06:02 & 2459905.75173-2459905.91849 & 4.0 & 229 & L2.3\\ 
1.1 m Hall & 2022-11-22T06:35 & 2459905.77496-2459905.85868 & 2.0 & 39 & L2.3\\ 
1.1 m Hall & 2022-11-23T06:29 & 2459906.77020-2459907.04413 & 6.6 & 123 & L2.3\\ 
\hline
1.5 m Danish & 2022-11-24T05:12 & 2459907.71703-2459907.85243 & 3.2 & 77 & L2.4\\ 
1.1 m Hall & 2022-11-24T06:24 & 2459907.76714-2459908.03644 & 6.5 & 111 & L2.4\\ 
0.9 m Spacewatch & 2022-11-24T07:43 & 2459907.82161-2459908.04074 & 5.3 & 124 & L2.4\\ 
1.5 m AZT-22 & 2022-11-24T20:49 & 2459908.36785-2459908.44936 & 2.0 & 76 & L2.4\\ 
1 m LCOGT & 2022-11-25T01:01 & 2459908.54258-2459908.76786 & 5.4 & 166 & L2.4\\ 
1.1 m Hall & 2022-11-25T06:30 & 2459908.77131-2459909.04076 & 6.5 & 117 & L2.4\\ 
1.5 m Danish & 2022-11-25T06:39 & 2459908.77729-2459908.85502 & 1.9 & 47 & L2.4\\ 
0.9 m Spacewatch & 2022-11-25T08:28 & 2459908.85347-2459909.03805 & 4.4 & 99 & L2.4\\ 
\hline
1.5 m Danish & 2022-11-26T05:56 & 2459909.74773-2459909.85536 & 2.6 & 47 & L2.5\\ 
1.1 m Hall & 2022-11-26T06:14 & 2459909.76004-2459910.04018 & 6.7 & 100 & L2.5\\ 
1.5 m Danish & 2022-11-27T05:36 & 2459910.73347-2459910.85399 & 2.9 & 61 & L2.5\\ 
1.1 m Hall & 2022-11-27T06:09 & 2459910.75637-2459911.04136 & 6.8 & 125 & L2.5\\ 
1.5 m AZT-22 & 2022-11-27T19:24 & 2459911.30869-2459911.55133 & 5.8 & 160 & L2.5\\ 
\hline
0.6 m Ondrejov & 2022-11-27T22:02 & 2459911.41807-2459911.53662 & 2.8 & 44 & L2.6\\ 
1.5 m Danish & 2022-11-28T05:36 & 2459911.73380-2459911.78206 & 1.2 & 28 & L2.6\\ 
0.6 m Stara Lesna & 2022-11-28T23:27 & 2459912.47749-2459912.63264 & 3.7 & 58 & L2.6\\ 
1.1 m Hall & 2022-11-29T06:05 & 2459912.75399-2459913.01913 & 6.4 & 117 & L2.6\\ 
1 m LCOGT & 2022-11-30T00:51 & 2459913.53587-2459913.76175 & 5.4 & 162 & L2.6\\ 
1.5 m Danish & 2022-11-30T04:54 & 2459913.70480-2459913.85738 & 3.7 & 78 & L2.6\\ 
1.1 m Hall & 2022-11-30T06:03 & 2459913.75269-2459914.04306 & 7.0 & 114 & L2.6\\ 
2.4 m MRO & 2022-11-30T06:19 & 2459913.76355-2459913.99664 & 5.6 & 353 & L2.6\\ 
0.9 m Spacewatch & 2022-11-30T06:54 & 2459913.78763-2459914.03500 & 5.9 & 133 & L2.6\\ 
1 m LCOGT & 2022-11-30T06:50 & 2459913.78537-2459913.90697 & 2.9 & 89 & L2.6\\ 
\hline
1 m LCOGT & 2022-12-01T02:46 & 2459914.61557-2459914.74010 & 3.0 & 88 & L2.7\\ 
1.1 m Hall & 2022-12-01T05:48 & 2459914.74174-2459914.93801 & 4.7 & 82 & L2.7\\ 
1.5 m Danish & 2022-12-01T06:06 & 2459914.75456-2459914.85574 & 2.4 & 51 & L2.7\\ 
2.4 m MRO & 2022-12-01T06:15 & 2459914.76075-2459914.97346 & 5.1 & 240 & L2.7\\ 
0.9 m Spacewatch & 2022-12-01T07:13 & 2459914.80119-2459915.02296 & 5.3 & 142 & L2.7\\ 
0.5 m Sugarloaf & 2022-12-02T05:32 & 2459915.73117-2459915.94429 & 5.1 & 44 & L2.7\\ 
\hline
2 m Faulkes-N & 2022-12-14T10:02 & 2459927.91809-2459928.13465 & 5.2 & 249 & L3.1\\ 
4.3 m LDT & 2022-12-15T07:06 & 2459928.79595-2459929.06886 & 6.5 & 488 & L3.1\\ 
\hline
1 m LCOGT & 2022-12-16T23:41 & 2459930.48726-2459930.65677 & 4.1 & 116 & L3.2\\ 
1.5 m Danish & 2022-12-17T04:07 & 2459930.67167-2459930.85421 & 4.4 & 64 & L3.2\\ 
1.1 m Hall & 2022-12-17T04:23 & 2459930.68277-2459931.02634 & 8.2 & 120 & L3.2\\ 
\hline
0.6 m TN & 2022-12-18T22:18 & 2459932.42977-2459932.72830 & 7.2 & 139 & L3.3\\ 
1.5 m Danish & 2022-12-19T03:42 & 2459932.65464-2459932.85511 & 4.8 & 120 & L3.3\\ 
1.1 m Hall & 2022-12-19T04:10 & 2459932.67420-2459933.00020 & 7.8 & 128 & L3.3\\ 
1 m Tian-Shan & 2022-12-19T16:09 & 2459933.17361-2459933.43726 & 6.3 & 148 & L3.3\\ 
1.5 m Danish & 2022-12-20T03:54 & 2459933.66318-2459933.82730 & 3.9 & 79 & L3.3\\ 
0.8 m IAC80 & 2022-12-20T04:05 & 2459933.67079-2459933.79256 & 2.9 & 46 & L3.3\\ 
1.1 m Hall & 2022-12-20T04:51 & 2459933.70215-2459933.83654 & 3.2 & 58 & L3.3\\ 
0.6 m TN & 2022-12-21T00:01 & 2459934.50107-2459934.64302 & 3.4 & 73 & L3.3\\ 
1.1 m Hall & 2022-12-21T04:15 & 2459934.67758-2459934.99702 & 7.7 & 117 & L3.3\\ 
0.5 m Sugarloaf & 2022-12-21T04:06 & 2459934.67103-2459934.95380 & 6.8 & 127 & L3.3\\ 
\hline
0.7 m AC-32 & 2022-12-21T18:51 & 2459935.28582-2459935.60357 & 7.6 & 120 & L3.4\\ 
1.5 m Danish & 2022-12-22T03:31 & 2459935.64693-2459935.85438 & 5.0 & 115 & L3.4\\ 
1.5 m Danish & 2022-12-23T04:03 & 2459936.66886-2459936.82079 & 3.6 & 86 & L3.4\\ 
0.7 m AC-32 & 2022-12-23T17:49 & 2459937.24273-2459937.56234 & 7.7 & 127 & L3.4\\ 
\hline
1.8 m BOAO & 2022-12-24T14:05 & 2459938.08730-2459938.37919 & 7.0 & 129 & L3.5\\ 
2.4 m MRO & 2022-12-25T03:37 & 2459938.65106-2459938.98469 & 8.0 & 465 & L3.5\\ 
1.1 m Hall & 2022-12-25T04:20 & 2459938.68114-2459938.77319 & 2.2 & 39 & L3.5\\ 
1.1 m Hall & 2022-12-26T04:46 & 2459939.69905-2459939.99006 & 7.0 & 124 & L3.5\\ 
0.5 m Sugarloaf & 2022-12-25T03:36 & 2459938.65060-2459938.94060 & 7.0 & 32 & L3.5\\ 
0.5 m Sugarloaf & 2022-12-26T04:02 & 2459939.66866-2459939.94683 & 6.7 & 96 & L3.5\\ 
1.8 m BOAO & 2022-12-26T13:04 & 2459940.04469-2459940.36915 & 7.8 & 160 & L3.5\\ 
\hline
0.6 m Ondrejov & 2022-12-27T22:33 & 2459941.43962-2459941.62416 & 4.4 & 73 & L3.6\\ 
1.5 m Danish & 2022-12-30T02:53 & 2459943.62060-2459943.85782 & 5.7 & 141 & L3.6\\ 
1.5 m Danish & 2022-12-31T04:15 & 2459944.67764-2459944.85225 & 4.2 & 103 & L3.6\\ 
1 m LCOGT & 2022-12-31T06:16 & 2459944.76126-2459944.89522 & 3.2 & 98 & L3.6\\ 
\hline
1.5 m Danish & 2023-01-11T01:50 & 2459955.57649-2459955.81858 & 5.8 & 103 & L4.1\\ 
1 m LCOGT & 2023-01-14T05:46 & 2459958.74075-2459958.90387 & 3.9 & 67 & L4.1\\ 
0.6 m TN & 2023-01-15T00:13 & 2459959.50919-2459959.65642 & 3.5 & 73 & L4.1\\ 
1.5 m Danish & 2023-01-15T01:32 & 2459959.56451-2459959.80226 & 5.7 & 96 & L4.1\\ 
1 m LCOGT & 2023-01-16T07:23 & 2459960.80828-2459960.94995 & 3.4 & 64 & L4.1\\ 
1.5 m Danish & 2023-01-18T01:26 & 2459962.55987-2459962.79337 & 5.6 & 98 & L4.1\\ 
\hline
1.5 m Danish & 2023-01-24T01:02 & 2459968.54363-2459968.69672 & 3.7 & 55 & L4.2\\ 
1 m LCOGT & 2023-01-24T20:06 & 2459969.33776-2459969.58969 & 6.0 & 102 & L4.2\\ 
4.3 m LDT & 2023-01-25T02:08 & 2459969.58954-2459970.01919 & 10.3 & 218 & L4.2\\ 
1 m LCOGT & 2023-01-26T01:57 & 2459970.58135-2459970.80766 & 5.4 & 93 & L4.2\\ 
1 m LCOGT & 2023-01-26T21:19 & 2459971.38821-2459971.55048 & 3.9 & 72 & L4.2\\ 
1.5 m Danish & 2023-01-28T01:01 & 2459972.54294-2459972.76243 & 5.3 & 73 & L4.2\\ 
2.4 m MRO & 2023-01-28T01:45 & 2459972.57311-2459972.75924 & 4.5 & 169 & L4.2\\ 
1.5 m Danish & 2023-01-29T01:30 & 2459973.56271-2459973.76053 & 4.7 & 80 & L4.2\\ 
2.4 m MRO & 2023-01-29T02:19 & 2459973.59682-2459973.74686 & 3.6 & 116 & L4.2\\ 
2.4 m MRO & 2023-01-30T02:24 & 2459974.60024-2459974.74168 & 3.4 & 77 & L4.2\\ 
\hline
2.4 m MRO & 2023-02-11T01:42 & 2459986.57115-2459986.90646 & 8.0 & 255 & L5.1\\
\hline
4.3 m LDT & 2023-02-17T02:32 & 2459992.60565-2459992.93700 & 8.0 & 80 & L5.2\\ 
2.4 m MRO & 2023-02-17T07:06 & 2459992.79600-2459992.93543 & 3.3 & 44 & L5.2\\ 
4.3 m LDT & 2023-02-21T02:15 & 2459996.59395-2459996.85968 & 6.4 & 117 & L5.2\\ 
2.4 m MRO & 2023-02-25T03:21 & 2460000.63968-2460000.82854 & 4.5 & 255 & L5.2\\ 
\enddata
\end{deluxetable*}

\section{Supplemental Material}

All of the lightcurve measurements and associated decompositions are included here as a supplemental data file. The columns in this file are: light time corrected Julian Date, measured apparent magnitude, differential magnitudes, decomposed residuals, and IDs for the observing runs and decompositions.

\bibliography{didymos_phot}{}
\bibliographystyle{aasjournal}



\end{document}